\documentclass[12pt]{article}
\usepackage{amsmath,amssymb}
\usepackage{bm}
\usepackage{color}
\usepackage{graphicx}
\usepackage{cite}
\usepackage{mathtools}
\usepackage{breqn}

\usepackage{ulem}

\usepackage{slashed}

\usepackage{here}

\usepackage{comment}

\begin{document}

\title{
\begin{flushright}
\ \\*[-80pt]
\begin{minipage}{0.2\linewidth}
\normalsize
EPHOU-22-003\\*[50pt]
\end{minipage}
\end{flushright}
{\Large \bf
Modular symmetry anomaly and non-perturbative neutrino mass terms in magnetized orbifold models
\\*[20pt]}}

\author{
Shota Kikuchi$^{a}$,
~Tatsuo Kobayashi$^{a}$, \\
 Kaito Nasu$^{a}$, 
~Hikaru Uchida$^{a}$, and 
~Shohei Uemura$^{b}$ 
\\*[20pt]
\centerline{
\begin{minipage}{\linewidth}
\begin{center}
{\it \normalsize
${}^{a}$Department of Physics, Hokkaido University, Sapporo 060-0810, Japan\\
${}^{b}$CORE of STEM, Nara Women's University, Nara 630-8506, Japan} \\*[5pt]
\end{center}
\end{minipage}}
\\*[50pt]}

\date{
\centerline{\small \bf Abstract}
\begin{minipage}{0.9\linewidth}
\medskip
\medskip
\small
We study the modular symmetry anomaly in magnetized orbifold models.
The non-perturbative effects such as D-brane instanton effects can break tree-level symmetry.
We study which part of the modular symmetry is broken explicitly by Majorana mass terms 
with three generations of neutrinos.
The modular weight of neutrino mass terms does not match with other coupling terms in 
tree-level Lagrangian.
In addition, the $Z_N$ symmetry of the modular flavor symmetry is broken and 
a certain normal subgroup of the modular flavor symmetry remains in neutrino mass terms.
\end{minipage}
}

\begin{titlepage}
\maketitle
\thispagestyle{empty}
\end{titlepage}

\newpage


\section{Introduction}
\label{Intro}

Superstring theory predicts extra  six-dimensional (6D) compact space in addition to our four-dimensional (4D) space-time.
Certain compactifications such as torus and some orbifold compactifications have a kind of geometrical symmetries, called as the modular symmetry.
Then, the modular symmetry appears in 
4D low energy effective field theory \cite{Ferrara:1989bc}.
Furthermore, the modular symmetry transforms zero-modes each other, e.g in heterotic orbifold models \cite{Ferrara:1989qb,Lerche:1989cs,Lauer:1990tm} 
and magnetized D-brane models \cite{Kobayashi:2018rad,Kobayashi:2018bff,Ohki:2020bpo,Kikuchi:2020frp,Kikuchi:2020nxn,
Kikuchi:2021ogn,Almumin:2021fbk}.
(See also \cite{Baur:2019iai,Nilles:2020kgo,Baur:2020jwc,Nilles:2020gvu}.)
That is, the modular symmetry can include a flavor symmetry among three generations of quarks and leptons in particle physics
\footnote{Calabi-Yau compactifications have many moduli, and they 
have larger symplectic modular symmetries \cite{Strominger:1990pd,Candelas:1990pi,Ishiguro:2020nuf,Ishiguro:2021ccl}.}.

Inspired by the above aspects, recently the modular flavor symmetric models have been studied intensively 
in the bottom-up approach.
(See for early works \cite{Feruglio:2017spp}.)\footnote{See for more recent list references e.g. in Ref.~\cite{Kikuchi:2022txy}.}
Indeed, the modular symmetry includes $S_3, A_4, S_4$, and $A_5$ as finite modular groups \cite{deAdelhartToorop:2011re}, 
and these non-Abelian discrete symmetries 
are often used for model building for quark and lepton flavor models in the bottom-up approach~\cite{
	Altarelli:2010gt,Ishimori:2010au,Ishimori:2012zz,Hernandez:2012ra,
	King:2013eh,King:2014nza}.

Symmetries at tree level are broken by quantum effects, that is anomaly.
Then, symmetry breaking terms appear by non-perturbative instanton effects in field theory.
The modular symmetry can also be anomalous.
Indeed, the modular symmetry anomaly, which is relevant to the automorphy factor except the flavor symmetry, 
were studied in 4D low energy effective field theory derived from heterotic string theory \cite{Derendinger:1991hq,Ibanez:1992hc}.
Such anomalies can be canceled by the 4D Green-Schwarz mechanism due to the axionic shift of the dilaton multiplet.
That leads to important aspects.
The moduli and dilaton mix in one-loop effective field theory.
Furthermore, mixed anomalies between the modular symmetry and gauge symmetries should be 
universal for all of the gauge symmetries in heterotic models.
This universality condition on mixed anomalies constrains massless spectra.
Several phenomenological applications were carried out, e.g. the gauge coupling unification, Yukawa couplings, 
and the hidden sector \cite{Ibanez:1992hc,Kobayashi:1993vn,Kawabe:1994mj}.
Moreover, the modular symmetry anomaly relevant to the automorphy factor was studied 
in intersecting and magnetized D-brane models \cite{Kobayashi:2016ovu}.
Similar to heterotic models, moduli mixing in one-loop effective field theory is required to
cancel the anomaly by the 4D Green-Schwarz mechanism.

In addition to the automorphy factor, anomalies of the modular flavor symmetries were studied 
in 4D effective field theory of magnetized D-brane models \cite{Kariyazono:2019ehj}.
Anomalous subsymmetries in the modular flavor symmetry correspond to  discrete symmetries of $U(1)$ gauge groups.
Thus, those anomalies can be canceled by the same Green-Schwarz mechanism to cancel the $U(1)$ anomalies.

Anomalies of the flavor symmetries are important.
If those are anomalous, the tree-level flavor symmetries are not exact at quantum level, 
but symmetry breaking terms appear non-perturbatively and affect the flavor structure.
In this paper, we study more about anomalies of the modular symmetries 
and non-perturbative symmetry breaking terms in 4D effective field theory of magnetized D-brane models.

Magnetized D-brane models lead to quite interesting low energy effective field theory~\cite{Bachas:1995ik,Blumenhagen:2000wh,Angelantonj:2000hi,Blumenhagen:2000ea,Cremades:2004wa,Abe:2008fi,Abe:2013bca,Abe:2014noa,Kobayashi:2017dyu,Sakamoto:2020pev,Abe:2008sx,Abe:2015yva,Hoshiya:2020hki}.
Yukawa couplings as well as higher dimensional couplings can be calculated by overlap integrations of wavefunctions on the compact space~\cite{Cremades:2004wa,Abe:2009dr,Fujimoto:2016zjs}.
Actually, from such compactifications, realistic quark masses and mixing angles as well as charged lepton masses have been studied~\cite{Abe:2012fj,Abe:2014vza,Fujimoto:2016zjs,Kobayashi:2016qag}.
In D-brane models, D-brane instanton effects induce new terms such as right-handed Majorana neutrino mass terms~\cite{Blumenhagen:2006xt,Ibanez:2006da,Cvetic:2007ku}, and explicit forms were also studied in magnetized models~\cite{Kobayashi:2015siy,Hoshiya:2021nux}.
We study the modular symmetry anomaly and breaking symmetries due to 
neutrino mass terms induced by D-brane instanton effects.
Such studies have implications on 4D modular flavor symmetric models.

This paper is organized as follows.
In section~\ref{sec:modular-symm-anomaly}, we briefly review the modular symmetry and its anomaly.
In section~\ref{sec:Revew}, we review Majorana neutrino masses generated by D-brane instanton effects in magnetized orbifold models.
In section~\ref{sec:Modularanomaly}, we study modular symmetry anomaly of the Majorana mass 
terms, generally.
In particular, in section~\ref{sec:Modularflavoranomaly}, we study modular flavor symmetry anomalies of Majorana mass terms for four types of three generations of right-handed neutrinos with modular symmetry on magnetized $T^2/Z_2$ orbifold, explicitly.
In section~\ref{sec:more}, we discuss more on possible corrections due to D-brane instanton effects.
We conclude this study in section~\ref{sec:conclusion}.


\section{Modular symmetry and anomaly}
\label{sec:modular-symm-anomaly}

In this section, we give a brief review on the modular symmetry and its anomalies.

\subsection{Modular symmetry}
\label{sec:mod-symm}

The modular group $\Gamma \equiv SL(2,\mathbb{Z})$ is the group of $(2\times 2)$ matrices,
\begin{align}
\gamma=
\begin{pmatrix}
a & b \\
c & d
\end{pmatrix},
\end{align}
where  $a,b,c,d$ are integers satisfying $ad-bc=1$.
The generators of $\Gamma$ are given by
\begin{align}
S =
\begin{pmatrix}
0 & 1 \\
-1 & 0
\end{pmatrix},
\quad
T =
\begin{pmatrix}
1 & 1 \\
0 & 1
\end{pmatrix},
\label{eq:ST}
\end{align}
and they satisfy the following algebraic relations:
\begin{align}
\begin{array}{l}
S^2 = - \mathbb{I}, \qquad 
S^4=(ST)^3=\mathbb{I},
\end{array}
\label{eq:algGamma}
\end{align}
where $\mathbb{I}$ denotes the unit matrix.

Under the modular symmetry, the modulus $\tau$ transforms as
\begin{align}
\tau ~\to~ \gamma \tau =  \frac{a\tau + b}{c\tau + d}.
\end{align}
The modular forms are holomorphic functions of $\tau$, which transform 
as 
\begin{align}
f_i(\tau) ~\to~f_i(\gamma\tau)=J_k(\gamma,\tau)\rho_{ij}(\gamma) f_j(\tau),
\end{align}
where $J_k(\gamma,\tau)=(c\tau+d)^k$ denotes the automorphy factor with the modular weight $k$ and 
$\rho_{ij}(\gamma)$ is unitary matrix.
The modular forms transform as 
\begin{align}
f_i(\tau) ~\to~f_i(\gamma'\tau)=J_k(\gamma',\tau)f_i(\tau),
\end{align}
for $\gamma'$ in a certain subgroup such as congruence subgroups.

\subsection{Matter fields and anomalies}
\label{sec:anomaly}

Here, we give a brief review on the modular symmetry anomaly 
in string-derived low energy effective field theory.
In 4D effective field theory, chiral matter fields $\phi_i$ have the following 
K\"{a}hler metric,
\begin{align}
\frac{1}{(2{{\rm Im}\tau)^{k_i}}} |\phi_i|^2, \label{eq:Kahlermetric}
\end{align}
and also transform \cite{Ferrara:1989bc}
\begin{align}
\phi_i ~\to~J_{-{k_i}}(\gamma,\tau)\rho_{ij}(\gamma) \phi_j,
\end{align}
under the modular symmetry.
The matrix $\rho_{ij}(\gamma)$ represents the flavor symmetry.

In Refs.~\cite{Derendinger:1991hq,Ibanez:1992hc}, 
the modular symmetry anomalies, which are relevant to the automorphy factor $J_{-{k_i}}(\gamma,\tau)$, 
were studied, i.e. for $\gamma'$ satisfying $\rho_{ij}(\gamma')=\delta_{ij}$.
The anomaly coefficients of mixed anomalies with the $G_a$ gauge symmetry are written by \cite{Derendinger:1991hq},
\begin{align}
A_a=-C(G_a) + \sum_{i}T(R^i_a) (1+2k_i),
\end{align}
where $C(G_a)$ is the quadratic Casimir of $G_a$ and $T(R^i_a)$ denotes the 
Dynkin index of the representation $R^i_a$ of chiral matter field $\phi_i$ under $G_a$.

This anomaly can be canceled by the 4D Green-Schwarz mechanism, where 
other moduli $T_\alpha$ in the gauge kinetic function $f_a(T_\alpha)$ of the gauge group $G_a$
 transform under the modular symmetry \cite{Derendinger:1991hq,Ibanez:1992hc},
\begin{align}
T_\alpha~\to~T_\alpha + \frac{1}{8\pi^2} \delta^\alpha_{GS} \ln (c\tau +d).
\label{eq:GS}
\end{align}
One-loop correction on the gauge kinetic function, which depends on $\tau$, may also contribute partly to 
the anomaly cancellation.
In heterotic string theory on orbifolds, the dilaton corresponds to the Green-Schwarz field.

The tree-level K\"ahler potential of moduli, 
\begin{align}
-\ln(2{\rm Im}\tau)-\sum_\alpha \ln (T_\alpha + \bar T_\alpha),
\end{align}
is not invariant under the modular symmetry because of the above transformation (\ref{eq:GS}).
The modular invariant K\"ahler potential is written by 
\begin{align}
-\ln(2{\rm Im}\tau)-\sum_\alpha \ln (T_\alpha + \bar T_\alpha + \frac{1}{8\pi^2} \delta^\alpha_{GS} \ln {\rm Im} \tau).
\end{align}
Thus, at this level, the modulus $\tau$ and other moduli $T_\alpha$ mix each other 
in the K\"ahler potential.
This study was extended to 4D low energy effective field theory derived from 
intersecting and magnetized D-brane models \cite{Kobayashi:2016ovu}.

Furthermore, the anomalies corresponding to $\rho_{ij}(\gamma)$ were studied in Ref.~\cite{Kariyazono:2019ehj}.
The matrix $\rho_{ij}(\gamma)$ represents the flavor symmetry corresponding to a non-Abelian discrete group.
In field theory with non-Abelain discrete group, 
the anomaly-free condition for the mixed anomaly with the non-Abelian gauge group $G_a$ 
is written by \cite{Araki:2008ek,Chen:2015aba,Kobayashi:2021xfs,Gripaios:2022vvc},
\begin{align}
({\rm det}\rho_{ij}(\gamma))^{\sum_i 2T_2(R\i_a)}=1 .
\end{align}
The subsymmetry corresponding to the element $\gamma$ with ${\rm det}\rho_{ij}(\gamma)=1$ is always 
anomaly free.
The other part of symmetry corresponding to the element with ${\rm det}\rho_{ij}(\gamma) \neq 1$ 
can be anomalous, although it depends on $\sum_i 2T_2(R\i_a)$.
Following this criteria, the anomalies were studied in Ref.~\cite{Kariyazono:2019ehj} in magnetized D-brane models.
It was found that the anomalous part can be embedded in discrete part of anomalous $U(1)$ gauge symmetry.

The $U(1)$ anomaly can be canceled by the 4D Green-Schwarz mechanism 
\cite{Witten:1984dg,Dine:1987xk,Lerche:1987sg,Kobayashi:1996pb,Ibanez:1998qp,Lalak:1999bk}, which 
requires the shift of the moduli $T_\alpha$ 
\begin{align}
T_\alpha~\to~T_\alpha + A^\alpha \Lambda,
\label{eq:GSU1}
\end{align}
under the gauge transformation of the $U(1)$ vector multiplet $V$,
\begin{align}
V~\to~V + \Lambda+\bar \Lambda,
\end{align}
where $\Lambda$ denotes the gauge transformation parameter.
Since the anomalous symmetries corresponding to ${\rm det}\rho_{ij}(\gamma) \neq 1$ 
is embedded in a discrete part of anomalous $U(1)$ gauge symmetry,  
anomalies of the modular flavor symmetries are also canceled by the same mechanism.

The flavor symmetry is quite important in low energy effective field theory.
In what follows, 
we study more about its anomalies.
Non-perturbative effects such as D-brane instanton effects break the tree-level symmetry and induce 
breaking terms in low energy effective field theory.
One of important terms in low energy effective field theory is right-handed Majorana neutrino mass terms, 
which can be induced by D-brane instanton effects.
In following sections, we study which part of modular flavor symmetry is broken by 
neutrino mass terms by D-brane instanton effects in magnetized orbifold models.

\section{Majorana neutrino mass terms in magnetized orbifold models}
\label{sec:Revew}

In this section, we review magnetized orbifold models and 
Majorana neutrino masses generated by D-brane instanton effects in magnetized orbifold models.


\subsection{Neutrinos in magnetized $T^2/Z_2$ orbifold compactifications}

First, we consider ${\cal M}_4 \times (T^2 \times X_4)/Z_2$ as 10D space-time in IIB superstring theory, 
where ${\cal M}_4$ is our 4D space-time and $X_4$ is a 4D compact space.
The action of $Z_2$  for $T^2$ is given by the $Z_2$ twist of the $T^2$ coordinate, thus it includes toroidal orbifold, $T^2/Z_2$.
We also introduce D-branes wrapping $p$-cycles on the compact space, $(T^2 \times X_4)/Z_2$.
The low energy effective theory of the open strings stretching between D-branes is given by supersymmetric gauge theory, and magnetic fluxes can be turned on.
Suppose that neutrinos $N_a$ correspond to zero-modes of open strings between two stacks of D-branes, D$_{N1}$ and  D$_{N2}$, with different quantized magnetic fluxes, denoted as $\overline{M}_{N1}$ and $\overline{M}_{N2}$, respectively.
For simplicity, we assume that they are D9-branes spreading the whole 10D space-time. 
We denote the difference of their magnetic fluxes on $T^2$ as $M_N \equiv \overline{M}_{N1} - \overline{M}_{N2}$, 
which appears in the zero-mode equation of neutrinos.
The generation number of the neutrinos, $N_a$, is determined by this $M_N$ as well as boundary conditions on $T^2/Z_2$ such as the $Z_2$ parity $m \in \{0,1\}$ and the Scherk-Schwarz (SS) phases $\alpha_1, \alpha_{\tau} \in \{0,1/2\}$ as will be shown.
Note that although the total generation number is also affected by degeneracy on $X_4$ and the $Z_2$ action on $X_4$,
we assume that the degeneracy on $X_4$ is just one and the action of $Z_2$ for the wavefunction on the $X_4$ is trivial,
hence the generation number is given by the degeneracy on $T^2/Z_2$.
In this case, the contribution from $X_4$ to the neutrino sector is just a flavor independent overall factor in Yukawa couplings and neutrino masses.
Hereafter, we concentrate on the 6D ${\cal M}_4 \times T^2/Z_2$,
where we denote the real coordinate of ${\cal M}_4$ and the complex coordinate of $T^2/Z_2$ as $x$ and $z$, respectively.

Now, we briefly review magnetized $T^2/Z_2$ orbifold compactifications~\cite{Abe:2008fi,Abe:2013bca}.
the $T^2/Z_2$ orbifold is constructed by the identification, $z \sim z+1 \sim z+\tau \sim -z$, where $\tau$ is the complex structure modulus of $T^2$ as well as $T^2/Z_2$.
A two-dimensional (2D) spinor on $T^2/Z_2$ with $U(1)$ unit charge $q=1$ under the magnetic flux $M$, the SS phases $(\alpha_1,\alpha_{\tau})$, and $Z_2$ parity $m$,
\begin{align}
\psi^{(\alpha_1,\alpha_{\tau};m), M}(z) = 
\begin{pmatrix}
\psi_+^{(\alpha_1,\alpha_{\tau};m), M}(z) \\
\psi_-^{(\alpha_1,\alpha_{\tau};m), M}(z)
\end{pmatrix},
\end{align}
should satisfy the following boundary conditions;
\begin{align}
\begin{array}{l}
\psi_{\pm}^{(\alpha_1,\alpha_{\tau};m), M}(z+1) =
e^{2\pi i \alpha_1} e^{\pi iM\frac{{\rm Im}z}{{\rm Im}\tau}} \psi_{\pm}^{(\alpha_1,\alpha_{\tau};m), M}(z), \\
\psi_{\pm}^{(\alpha_1,\alpha_{\tau};m), M}(z+\tau) =
e^{2\pi i \alpha_{\tau}} e^{\pi iM\frac{{\rm Im}(\bar{\tau}z)}{{\rm Im}\tau}} \psi_{\pm}^{(\alpha_1,\alpha_{\tau};m), M}(z), \\
\psi_{\pm}^{(\alpha_1,\alpha_{\tau};m), M}(-z) =
(-1)^m \psi_{\pm}^{(\alpha_1,\alpha_{\tau};m), M}(z).
\end{array}
\end{align} 
When $M$ is positive (negative), only $\psi_+$ ($\psi_-$) has zero-mode solutions of Dirac equation $i\slashed{D}\psi(z) = 0$,
where the covariant derivative includes background $U(1)$ gauge potential which induces the magnetic flux $M$.
Hereafter, we consider only the case with the positive magnetic flux  for simplicity and we omit the notation of the chirality, ``+".
The $a$ th zero-mode solution can be expressed as
\begin{align}
&\psi^{(a+\alpha_1,\alpha_{\tau};m),M}(z) = \notag \\
&{\cal N}^a e^{\pi iMz\frac{{\rm Im}z}{{\rm Im}\tau}} \left( e^{2\pi i\frac{(a+\alpha_1)\alpha_{\tau}}{M}} \vartheta
\begin{bmatrix}
\frac{a+\alpha_1}{M} \\ -\alpha_{\tau}
\end{bmatrix}
(Mz,M\tau)
+ (-1)^{m-2\alpha_{\tau}}
e^{2\pi i \frac{M-(a+\alpha_1)}{M}} \vartheta
\begin{bmatrix}
\frac{M-(a+\alpha_1)}{M} \\ -\alpha_{\tau}
\end{bmatrix}
(Mz,M\tau) \right), \label{eq:wav} \\
&{\cal N}^a = \left\{
\begin{array}{l}
\frac{1}{2} \left(\frac{M}{{\cal A}^2}\right)^{\frac{1}{4}} \quad (a+\alpha_1=0,|M|/2) \\
\frac{1}{\sqrt{2}} \left(\frac{M}{{\cal A}^2}\right)^{\frac{1}{4}} \quad ({\rm otherwise})
\end{array}
\right.,
\end{align}
where ${\cal A}$ denotes the area of $T^2$
and $\vartheta$ denotes the Jacobi theta function given by
\begin{align}
\vartheta
\begin{bmatrix}
a\\
b
\end{bmatrix}
(\nu, \tau)
=
\sum_{l\in \mathbb{Z}}
e^{\pi i (a+l)^2\tau}
e^{2\pi i (a+l)(\nu+b)}.
\end{align}
The normalization factor, ${\cal N}^a$, is determined by
\begin{align}
\int_{T^2/Z_2} dzd\bar{z} \psi_+^{(a+\alpha_1,\alpha_{\tau};m),M}(z) \left(\psi_+^{(a+\alpha'_1,\alpha_{\tau};m),M}(z)\right)^{\ast} =
(2{\rm Im}\tau)^{-\frac{1}{2}} \delta_{a,a'}.
\label{eq:normalization}
\end{align}
Then, the number of zero-modes is shown in Table~\ref{tab:zeromode}.
Therefore, we can obtain such numbers of chiral fermions from the magnetized $T^2/Z_2$ orbifold.
The $a$ th generation of the neutrinos in the 4D space-time, $N_a(x)$, comes from the $a$ th zero-mode on the magnetized $T^2/Z_2$, $\psi^{(a+\alpha_1,\alpha_{\tau};m)_N, M_N}(z)$.
In particular, in the following calculations, we study models with three generations of  neutrinos.
\begin{table}[H]
\centering
\begin{tabular}{|c|c|c|} \hline
$(\alpha_1,\alpha_{\tau};m)$ & $M \in 2\mathbb{Z}$ & $M \in 2\mathbb{Z}+1$ \\ \hline \hline
$(0,0;0)$ & $\frac{M}{2}+1$ & $\frac{M+1}{2}$ \\ \hline
$(0,0;1)$ & $\frac{M}{2}-1$ & $\frac{M-1}{2}$ \\ \hline \hline
$(1/2,0;0)$ & $\frac{M}{2}$ & $\frac{M+1}{2}$ \\ \hline
$(1/2,0;1)$ & $\frac{M}{2}$ & $\frac{M-1}{2}$ \\ \hline \hline
$(0,1/2;0)$ & $\frac{M}{2}$ & $\frac{M+1}{2}$ \\ \hline
$(0,1/2;1)$ & $\frac{M}{2}$ & $\frac{M-1}{2}$ \\ \hline \hline
$(1/2,1/2;0)$ & $\frac{M}{2}$ & $\frac{M-1}{2}$ \\ \hline
$(1/2.1/2;1)$ & $\frac{M}{2}$ & $\frac{M+1}{2}$ \\ \hline
\end{tabular}
\caption{The number of zero-modes}
\label{tab:zeromode}
\end{table}


\subsection{Majorana neutrino masses induced by D-brane instanton effects}

Next, let us review the flavor structure of the Majorana neutrino masses generated by D-brane instanton effects in the magnetized orbifold models~\cite{Hoshiya:2021nux}.
The D-brane instanton is an instanton-like solution of string theory.
It is localized at a point in 4D space-time, but wrapping a cycle on the compact space.
When a D-brane instanton D$_{inst}$ with a magnetic flux $\overline{M}_{inst}$ exists, zero-modes $\beta_i$ ($\gamma_j$) appear between D$_{N1}$ (D$_{N2}$) and D$_{inst}$.
We denote the difference of their magnetic fluxes on $T^2/Z_2$ as $M_{\beta} \equiv \overline{M}_{N1} - \overline{M}_{inst}$ ($M_{\gamma} \equiv \overline{M}_{N2} - \overline{M}_{inst}$), which appears in the zero-mode equation of $\beta_i$ ($\gamma_j$).
The number of the zero-modes is determined by the magnetic flux $M_{\beta}$ ($M_{\gamma}$) as well as boundary conditions on $T^2/Z_2$: the $Z_2$ parity and the SS phases.
Hereafter, we consider that the $i$ th instanton zero-mode in the 4D space-time, $\beta_i(x)$ ($\gamma_i(x)$), which is localized at the point $x$ in 4D space-time, comes from the $i$ th  zero-mode on the magnetized $T^2/Z_2$, $\psi^{(i+\alpha_1,\alpha_{\tau};m)_{\beta}, M_{\beta}}(z)$ ($\psi^{(i+\alpha_1,\alpha_{\tau};m)_{\gamma}, M_{\gamma}}(z)$).

We give a comment on $U(1)$ gauge symmetries.
Each D-brane has a $U(1)$ gauge symmetry.
That is, D$_{N1}$ and D$_{N2}$ have $U(1)_1$ and $U(1)_2$ gauge symmetries, respectively.
Then, neutrinos have $(1,-1)$ charges under $U(1)_1\times U(1)_2$, while $\beta_i$ and $\gamma_i$ 
would have $(-1,0)$ and $(0,1)$ charges, respectively.
Both of $U(1)_1\times U(1)_2$ or their linear combination can be anomalous.
Such anomalies could be canceled by the 4D Green-Schwarz mechanism as will be shown.

There appear three point couplings of their zero-modes and neutrinos, 
\begin{align}
d_a^{ij} \beta_i(x)\gamma_j(x)N_a(x),
\end{align}
 where $d_a^{ij}$ denotes the coupling coefficients.
Due to the three point couplings, Majorana neutrino mass terms, $M_{ab}N_a(x)N_b(x)$, can be induced~\cite{Blumenhagen:2006xt,Ibanez:2006da} as
\begin{align}
M_{ab}N_a(x)N_b(x)
&= e^{-S_{cl}(T_\alpha,\overline{M}_{inst})} \int d^2\beta d^2\gamma e^{-d^{ij}_a\beta_i(x) \gamma_j(x) N_a(x)} \notag \\
&= e^{-S_{cl}(T_\alpha,\overline{M}_{inst})} (\varepsilon_{ij}\varepsilon_{k\ell} d^{ik}_{a}d^{j\ell}_b) N_a(x)N_b(x) \label{eq:neutrinomass} \\
&= e^{-S_{cl}(T_\alpha,\overline{M}_{inst})} m_{ab}N_a(x)N_b(x). \notag
\end{align}
Here, we give several comments in order.
First, $S_{cl}(T_\alpha,\overline{M}_{inst})$ denotes the classical action of the D-brane instanton which depends on 
the moduli $T_\alpha$ through 
D-brane instanton volume and magnetic flux in the compact space.
Note that there appears the axion of $T_\alpha$,  
to which the D-brane instanton couples in the imaginary part of $S_{cl}(T_\alpha,\overline{M}_{inst})$.
Second, both $\beta_i(x)$ and $\gamma_j(x)$ are two numbers of Grassmann zero-modes $(i,j=1,2)$.
Here, the Grassmann integration for the Grassmann field $\psi \ (\psi = \beta_i, \gamma_j)$ satisfies
\begin{align}
\int d\psi \psi = 1. \label{eq:Grassmannintegral}
\end{align}
Thus, the Majorana mass terms can be generated in the only case that the numbers of both zero-modes, $\beta_i(x)$ and $\gamma_j(x)$, are two.
In order to obtain Majorana masses of three neutrinos from three point couplings, $d_a^{ij} \beta_i(x)\gamma_j(x)N_a(x)$, their magnetic fluxes and SS phases as well as $Z_2$ parities should satisfy
\begin{align} 
\label{eq:zero-mode-condition}
&M_N = M_{\beta}+M_{\gamma}, \\ &(\alpha_1, \alpha_{\tau}; m)_N  \equiv (\alpha_1,\alpha_{\tau}; m)_{\beta}+(\alpha_1,\alpha_{\tau}; m)_{\gamma}\ ({\rm mod}\ 1).   \notag
\end{align} 
Otherwise, the  coupling vanishes,  $d_a^{ij} =0$.
When the above condition is satisfied, the coupling coefficients $d_a^{ij}$ can be calculated from
\begin{align}
d_a^{ij} =
\int_{T^2/Z_2} dzd\bar{z}  \psi^{(i+\alpha_1,\alpha_{\tau};m)_{\beta}, M_{\beta}}(z) \psi^{(j+\alpha_1,\alpha_{\tau};m)_{\gamma}, M_{\gamma}}(z) \left( \psi^{(a+\alpha_1,\alpha_{\tau};m)_N, M_N}(z) \right)^{\ast}.
\label{eq:dmatrix}
\end{align}
This comes from the following decomposition of 6D fields, $\beta^{6D}$, $\gamma^{6D}$, and $N^{6D}$;
\begin{align}
\begin{array}{l}
\beta^{6D} = \sum_{i} \beta_i^{(M;\alpha_1,\alpha_{\tau};m)_{\beta}}(x) \otimes \psi^{(i+\alpha_1,\alpha_{\tau};m)_{\beta}, M_{\beta}}(z), \\
\gamma^{6D} = \sum_{j} \gamma_j^{(M;\alpha_1,\alpha_{\tau};m)_{\gamma}}(x) \otimes \psi^{(j+\alpha_1,\alpha_{\tau};m)_{\gamma}, M_{\gamma}}(z), \\
N^{6D} = \sum_{a} N_a^{(M;\alpha_1,\alpha_{\tau};m)_{N}}(x) \otimes \left( \psi^{(a+\alpha_1,\alpha_{\tau};m)_N, M_N}(z) \right)^{\ast}.
\label{eq:decomposition}
\end{array}
\end{align}
All of two-number of instanton zero-modes, $\beta_i^{(M;\alpha_1,\alpha_{\tau};m)_{\beta}}$ and $\gamma_j^{(M;\alpha_1,\alpha_{\tau};m)_{\gamma}}$, which satisfy 
the above condition (\ref{eq:zero-mode-condition}), can contribute to generating the Majorana neutrino masses.
Thus, the total Majorana mass terms can be written as
\begin{align}
&M_{ab}N_a^{(M;\alpha_1,\alpha_{\tau};m)_{N}}(x)N_b^{(M;\alpha_1,\alpha_{\tau};m)_{N}}(x) = \notag \\
&\left( \sum_{M_{inst}} e^{-S_{cl}(T_\alpha,\overline{M}_{inst}^{M_{inst}})} \sum_{{\bm \alpha}_{inst}} m_{ab}^{(M,{\bm \alpha})_{inst}} \right) N_a^{(M;\alpha_1,\alpha_{\tau};m)_{N}}(x)N_b^{(M;\alpha_1,\alpha_{\tau};m)_{N}}(x),
\end{align} 
where we denote ${\bm \alpha}_X \equiv (\alpha_1, \alpha_{\tau})_X$, ${\bm \alpha}_{inst} \equiv ({\bm \alpha}_{\beta}, {\bm \alpha}_{\gamma})$, and $M_{inst} = (M_{\beta}, M_{\gamma})$.
Hereafter, we denote $(\alpha_1,\alpha_{\tau}) = (0,0)$, $(1/2,0)$, $(0,1/2)$, $(1/2,1/2)$ as $A$, $B$, $C$, $D$, for shortly, respectively.

In addition to these zero-modes, we have neutral zero-modes which correspond to the gauge multiplets on the D-brane instanton.
The number of the neutral zero-modes is also crucial since extra Grassmann integral can eliminate non-peruturbative superpotential.
These neutral zero-modes must be absorbed by interaction terms to obtain non-zero non-perturbative effects.
On the other hand, the integration of the neutral zero-modes does not affect the modular symmetry anomaly
since the wavefunction of gauge multiplets is constant on the compact space and does not transform under the modular group.
Thus, we investigate the flavor structure of non-perturbative Majorana mass term, 
assuming the integration of the neutral zero-modes is  properly absorbed by interaction terms in the present paper.


\section{Modular symmetry anomaly of Majorana neutrino mass terms}
\label{sec:Modularanomaly}

In this section, let us study modular symmetry anomaly of the Majorana neutrino mass terms in Eq.~(\ref{eq:neutrinomass}).
First, we briefly review the modular transformation of the wavefunctions and the coupling coefficients.
Under $\gamma 
\in \Gamma$ transformation, the modulus as well as the coordinate, $(z,\tau)$, transforms as
\begin{align}
\gamma : (z,\tau) \rightarrow (z',\tau') = \left( \frac{z}{c\tau+d}, \frac{a\tau+b}{c\tau+d} \right). \label{eq:modulartrans}
\end{align}
We call this transformation the modular transformation.
In particular, under $S$ and $T$ transformation defined in Eq.~(\ref{eq:ST}), $(z,\tau)$ transform as
\begin{align}
S : (z,\tau) \rightarrow (z',\tau') = \left( -\frac{z}{\tau}, -\frac{1}{\tau} \right), \quad
T : (z,\tau) \rightarrow (z',\tau') = \left( z, \tau+1 \right),
\label{eq:STztau}
\end{align}
respectively.
Thus, this modular transformation also satisfies Eq.~(\ref{eq:algGamma}).

Now, let us investigate the modular transformation for the wavefunctions on magnetized $T^2/Z_2$ in Eq.~(\ref{eq:wav})~\cite{Kikuchi:2020frp,Kikuchi:2021ogn}.
For this purpose, we introduce the double covering group of $\Gamma$, $\widetilde{\Gamma}$.
(See e.g Ref.~\cite{Liu:2020msy} and references therein.)
The generators of $\widetilde{\Gamma}$, $\widetilde{S}$ and $\widetilde{T}$, satisfy
\begin{align}
\begin{array}{l}
\widetilde{S}^2=\widetilde{Z}, \\
\widetilde{S}^4=(\widetilde{S}\widetilde{T})^3=\widetilde{Z}^2, \\
\widetilde{S}^8=(\widetilde{S}\widetilde{T})^6=\widetilde{Z}^4=\mathbb{I}, \\
\widetilde{Z}\widetilde{T} = \widetilde{T}\widetilde{Z},
\end{array}
\end{align}
where $\widetilde{Z}$ expands the center of $\Gamma$.
$\widetilde{S}$ and $\widetilde{T}$ transformations of $(z,\tau)$ are the same as $S$ and $T$ transformation in Eq.~(\ref{eq:STztau}).
Hence, under $\widetilde{\gamma} \in \widetilde{\Gamma}$ transformation, the wavefunctions in Eq.~(\ref{eq:wav}) transform as
\begin{align}
\widetilde{\gamma} ~:~ &\psi^{(a+\alpha_1,\alpha_{\tau};m),M}\left( z, \tau \right)~ \rightarrow \notag \\
&\psi^{(a+\alpha_1,\alpha_{\tau};m),M}\left( \frac{z}{c\tau+d}, \frac{a\tau+b}{c\tau+d} \right) =
\widetilde{J}_{1/2}(\widetilde{\gamma},\tau) \sum_{a'} \sum_{{\bm \alpha}'}
\rho(\widetilde{\gamma})_{aa'}^{{\bm \alpha}{\bm \alpha}'} \psi^{(a'+\alpha'_1,\alpha'_{\tau};m),M}\left( z, \tau \right),
\label{eq:modularwav}
\end{align}
where $\widetilde{J}_{1/2}$ and $\rho$ denote the automorphy factor with modular weight $1/2$ and the unitary matrix, respectively.
For $\widetilde{S}$ and $\widetilde{T}$ transformation, they can be expressed as
\begin{align}
&\widetilde{J}_{1/2}(\widetilde{S},\tau) = (-\tau)^{1/2}, \quad 
\rho(\widetilde{S})_{aa'}^{{\bm \alpha}{\bm \alpha}'} = \left\{
\begin{array}{ll}
{\cal N}^{a} {\cal N}^{a'} \frac{4e^{\pi i/4}}{\sqrt{M}} \cos \left( \frac{2\pi (a+\alpha_1)(a'+\alpha'_1)}{M}  \right) \delta_{(\alpha_{\tau},\alpha_1),(\alpha'_1,\alpha'_{\tau})} & (m=0) \\
{\cal N}^{a} {\cal N}^{a'} \frac{4ie^{\pi i/4}}{\sqrt{M}} \sin \left( \frac{2\pi (a+\alpha_1)(a'+\alpha'_1)}{M}  \right) \delta_{(\alpha_{\tau},\alpha_1),(\alpha'_1,\alpha'_{\tau})} & (m=1)
\end{array} \right., \\
&\widetilde{J}_{1/2}(\widetilde{T},\tau) = 1, \qquad 
\rho(\widetilde{T})_{aa'}^{{\bm \alpha}{\bm \alpha}'} = \left\{
\begin{array}{ll}
e^{\frac{\pi i(a+\alpha_1)^2}{M}} \delta_{a,a'} \delta_{(\alpha_1,\alpha_{\tau}-\alpha_1),(\alpha'_1,\alpha'_{\tau})} & (M \in 2\mathbb{Z}) \\
e^{\frac{\pi i(a+\alpha_1)^2}{M}} \delta_{a,a'} \delta_{(\alpha_1,\alpha_{\tau}-\alpha_1+\frac{1}{2}),(\alpha'_1,\alpha'_{\tau})} & (M \in 2\mathbb{Z}+1)
\end{array} \right.,
\label{eq:JrhoST}
\end{align}
respectively\footnote{Since the definition of wavefunctions in Eq.~(\ref{eq:wav}) is modified from ones in Ref.~\cite{Kikuchi:2021ogn}, the matrix forms are also modified from ones in Ref.~\cite{Kikuchi:2021ogn}.}.
Thus, under the modular transformation, in general, SS phases transform and then the fields such as $\beta^{(M;\alpha_1,\alpha_{\tau};m)_{\beta}}$ ($\gamma^{(M;\alpha_1,\alpha_{\tau};m)_{\gamma}}$) convert into other fields such as $\beta^{(M;\alpha'_1,\alpha'_{\tau};m)_{\beta}}$ ($\gamma^{(M;\alpha'_1,\alpha'_{\tau};m)_{\gamma}}$).
However, wavefunctions with $M \in 2\mathbb{Z}$ and $(\alpha_1,\alpha_{\tau})=(0,0)$ and ones with $M \in 2\mathbb{Z}+1$ and $(\alpha_1,\alpha_{\tau})=(1/2,1/2)$ are closed under the modular transformation.
Then, we consider the models, that the three generations of neutrinos come from such wavefunctions with the modular symmetry.
Namely, from the Table~\ref{tab:zeromode}, there are four cases with three generations of neutrinos: $(M; \alpha_1, \alpha_{\tau}; m)_N=(4;0,0;0)$, $(8;0,0;1)$, $(5;1/2,1/2;1)$, and $(7;1/2,1/2;0)$.
In these cases, the unitary matrices become unitary representations of $\widetilde{\Delta}(96) \simeq \Delta(48) \rtimes Z_8$, $\widetilde{\Delta}(384) \simeq \Delta(192) \rtimes Z_8$, $A_5 \times Z_8$, and $PSL(2,Z_7) \times Z_8$, respectively~\cite{Kikuchi:2021ogn}.

From Eqs.~(\ref{eq:decomposition}) and (\ref{eq:modularwav}), the 4D fields, $\beta_i(x)$, $\gamma_j(x)$, and $N_a(x)$, transform under the modular transformation as
\begin{align}
\begin{array}{l}
\widetilde{\gamma}~ :~ \beta_{i}^{(M;\alpha_1,\alpha_{\tau};m)_{\beta}}(x) ~\rightarrow~
\widetilde{J}_{-1/2}(\widetilde{\gamma},\tau) \sum_{i'} \sum_{{\bm \alpha}'_{\beta}}
\rho_{\beta}^{-1}(\widetilde{\gamma})_{ii'}^{{\bm \alpha}_{\beta}{\bm \alpha}'_{\beta}} \beta_{i'}^{(M;\alpha'_1,\alpha'_{\tau};m)_{\beta}}(x), \\
\widetilde{\gamma}~ :~ \gamma_{j}^{(M;\alpha_1,\alpha_{\tau};m)_{\gamma}}(x) ~\rightarrow~
\widetilde{J}_{-1/2}(\widetilde{\gamma},\tau) \sum_{j'} \sum_{{\bm \alpha}'_{\gamma}}
\rho_{\gamma}^{-1}(\widetilde{\gamma})_{jj'}^{{\bm \alpha}_{\gamma}{\bm \alpha}'_{\gamma}} \gamma_{j'}^{(M;\alpha'_1,\alpha'_{\tau};m)_{\gamma}}(x), \\
\widetilde{\gamma}~ :~ N_{a}^{(M;\alpha_1,\alpha_{\tau};m)_{N}}(x) ~\rightarrow~
\left( \widetilde{J}_{-1/2}(\widetilde{\gamma},\tau) \right)^{\ast} \sum_{a'} \rho_{N}^{T}(\widetilde{\gamma})_{aa'}^{{\bm \alpha}_{N}{\bm \alpha}_{N}} N_{a'}^{(M;\alpha_1,\alpha_{\tau};m)_{N}}(x),
\end{array}
\label{eq:4Dmodular}
\end{align}
respectively.
Note that Eq.~(\ref{eq:JrhoST}) satisfies $\rho(\widetilde{\gamma})^T = \rho(\widetilde{\gamma})$.
Therefore, in this case, the 4D three generations of  neutrinos, $N_a(x)$, transform non-trivially as triplets under the above discrete modular flavor transformation with the modular weight $-1/2$~\footnote{This is consistent with the
K\"{a}hler metric in Eq.~(\ref{eq:Kahlermetric}), obtained from Eq.~(\ref{eq:normalization})~~\cite{Kikuchi:2020frp,Kikuchi:2022txy}.}.
In the following section, we discuss their modular flavor symmetry anomalies individually.

On the other hand, from the modular transformation for wavefunctions in Eq.~(\ref{eq:modularwav}), we can find the modular transformation for three point coupling coefficients, $d_a^{ij,(M,{\bm \alpha})_{inst}}(\tau)$, in Eq.~(\ref{eq:dmatrix})~\cite{Hoshiya:2020hki};
\begin{align}
\widetilde{\gamma}~ :~ &d_{a}^{ij,(M,{\bm \alpha})_{inst}}(\tau) ~\rightarrow \notag \\
&d_{a}^{ij,(M,{\bm \alpha})_{inst}}\left( \frac{a\tau+b}{c\tau+d} \right) = 
\bigl|\widetilde{J}_{1/2}(\widetilde{\gamma},\tau) \bigl|^2 \widetilde{J}_{1/2}(\widetilde{\gamma},\tau)
\sum_{(i'j'a')} \sum_{{\bm \alpha}'_{inst}} 
\rho_d(\widetilde{\gamma})_{(ija)(i'j'a')}^{{\bm \alpha}_{inst}{\bm \alpha}'_{inst}}
d_{a'}^{i'j',(M,{\bm \alpha}')_{inst}}(\tau), \notag \\
&\rho_d(\widetilde{\gamma})_{(ija)(i'j'a')}^{{\bm \alpha}_{inst}{\bm \alpha}'_{inst}} = 
\rho_{\beta}(\widetilde{\gamma})_{ii'}^{{\bm \alpha}_{\beta}{\bm \alpha}'_{\beta}} \rho_{\gamma}(\widetilde{\gamma})_{jj'}^{{\bm \alpha}_{\gamma}{\bm \alpha}'_{\gamma}} \left( \rho_N(\widetilde{\gamma})_{aa'}^{{\bm \alpha}_N {\bm \alpha}_N} \right)^{\ast},
\label{eq:dmodular}
\end{align}
where the term, $\bigl|\widetilde{J}_{1/2}(\widetilde{\gamma},\tau) \bigl|^2=|c\tau+d|$, comes from the factor, $(2{\rm Im}\tau)^{-\frac{1}{2}}$, obtained by integration in Eq.~(\ref{eq:dmatrix}).
Thus, by combining Eqs.~(\ref{eq:4Dmodular}) and (\ref{eq:dmodular}), the three point coupling terms 
transform under the modular transformation as
\begin{align}
\widetilde{\gamma}~ :~ &d_{a}^{ij,(M,{\bm \alpha})_{inst}}(\tau)\beta_i^{(M;\alpha_1,\alpha_{\tau};m)_{\beta}}(x)\gamma_j^{(M;\alpha_1,\alpha_{\tau};m)_{\gamma}}(x)N_a^{(M;\alpha_1,\alpha_{\tau};m)_{N}}(x)
~\rightarrow \notag \\
&d_{a'}^{i'j',(M,{\bm \alpha}')_{inst}}(\tau) {\beta'}_{i'}^{(M;{\alpha'}_1,{\alpha'}_{\tau};m)_{\beta}}(x)\gamma_{j'}^{(M;{\alpha'}_1,{\alpha'}_{\tau};m)_{\gamma}}(x)N_{a'}^{(M;\alpha_1,\alpha_{\tau};m)_{N}}(x).
\label{eq:beta-gamma-trans}
\end{align}
In particular, if instanton zero-modes, $\beta$ and $\gamma$, also come from wavefunctions consistent with the modular symmetry, the above three point couplings are modular invariant.
Even if the above term transforms under the modular transformation, the total three point couplings which can generate Majorana masses, $\sum_{(M,{\bm \alpha})_{inst}} d_a^{ij,(M,{\bm \alpha})_{inst}}(\tau) \beta_i^{(M;\alpha_1,\alpha_{\tau};m)_{\beta}}(x)\gamma_j^{(M;\alpha_1,\alpha_{\tau};m)_{\gamma}}(x)N_a^{(M;\alpha_1,\alpha_{\tau};m)_{N}}(x)$, are modular invariant.

Now, let us see the modular transformation for the Majorana mass terms in Eq.~(\ref{eq:neutrinomass}).
Since we obtain the modular transformation for $d_a^{ij,(M,{\bm \alpha})_{inst}}$, we can find that the mass matrix elements, $m_{ab}^{(M,{\bm \alpha})_{inst}}(\tau)$, transform under the modular transformation as
\begin{align}
\widetilde{\gamma}~ :~ &m_{ab}^{(M,{\bm \alpha})_{inst}}(\tau)~ \rightarrow \notag \\
&m_{ab}^{(M,{\bm \alpha})_{inst}}\left( \frac{a\tau+b}{c\tau+d} \right) = 
\bigl|\widetilde{J}_{1}(\widetilde{\gamma},\tau) \bigl|^2 \widetilde{J}_{1}(\widetilde{\gamma},\tau)
\sum_{(a'b')} \sum_{{\bm \alpha}'_{inst}}
\rho_m(\widetilde{\gamma})_{(ab)(a'b')}^{{\bm \alpha}_{inst}{\bm \alpha}'_{inst}}
m_{a'b'}^{(M,{\bm \alpha}')_{inst}}(\tau), \notag \\
&\rho_m(\widetilde{\gamma})_{(ab)(a'b')}^{{\bm \alpha}_{inst}{\bm \alpha}'_{inst}} = 
{\rm det}[\rho_{inst}(\widetilde{\gamma})^{{\bm \alpha}_{inst}{\bm \alpha}'_{inst}}] \left( \rho_N(\widetilde{\gamma})_{aa'}^{{\bm \alpha}_N {\bm \alpha}_N} \right)^{\ast} \left( \rho_N(\widetilde{\gamma})_{bb'}^{{\bm \alpha}_N {\bm \alpha}_N} \right)^{\ast},
\label{eq:mmodular}
\end{align}
where we denote ${\rm det}[\rho_{inst}(\widetilde{\gamma})^{{\bm \alpha}_{inst}{\bm \alpha}'_{inst}}] \equiv {\rm det}[\rho_{\beta}(\widetilde{\gamma})^{{\bm \alpha}_{\beta}{\bm \alpha}'_{\beta}}] {\rm det}[\rho_{\gamma}(\widetilde{\gamma})^{{\bm \alpha}_{\gamma},{\bm \alpha}'_{\gamma}}]$.
Thus, by combining Eqs.~(\ref{eq:4Dmodular}) and (\ref{eq:mmodular}), the mass terms 
transform under the modular transformation as
\begin{align}
\widetilde{\gamma}~ :~ &m_{ab}^{(M,{\bm \alpha})_{inst}}(\tau) N_{a}^{(M;\alpha_1,\alpha_{\tau};m)_{N}}(x) N_{b}^{(M;\alpha_1,\alpha_{\tau};m)_{N}}(x)
~\rightarrow \notag \\
&\widetilde{J}_2(\widetilde{\gamma},\tau) {\rm det}[\rho_{inst}(\widetilde{\gamma})^{{\bm \alpha}_{inst}{\bm \alpha}'_{inst}}] m_{a'b'}^{(M,{\bm \alpha}')_{inst}}(\tau)^{{\bm \alpha}'_{inst}} N_{a'}^{(M;\alpha_1,\alpha_{\tau};m)_{N}}(x) N_{b'}^{(M;\alpha_1,\alpha_{\tau};m)_{N}}(x).
\label{eq:n-mass-anomaly}
\end{align}
This means that even if we consider instanton zero-modes consistent with the modular symmetry, the Majorana mass terms are generally not invariant under the modular transformation.
In other words, there appears modular symmetry anomaly, in general, in the Majorana mass terms generated by D-brane instanton effects.
Indeed, the anomalous factor, $\widetilde{J}_2(\widetilde{\gamma},\tau) {\rm det}[\rho_{inst}(\widetilde{\gamma})^{{\bm \alpha}_{inst}{\bm \alpha}'_{inst}}]$, comes from transformation for measures of instanton zero-modes in the path integral in Eq.~(\ref{eq:neutrinomass}), 
\begin{align}
\widetilde{\gamma}~ :~ &d^2\beta^{(M;\alpha_1,\alpha_{\tau};m)_{\beta}} d^2\gamma^{(M;\alpha_1,\alpha_{\tau};m)_{\gamma}}
~\rightarrow \notag \\
&\widetilde{J}_2(\widetilde{\gamma},\tau) {\rm det}[\rho_{inst}(\widetilde{\gamma})^{{\bm \alpha}_{inst}{\bm \alpha}'_{inst}}]
d^2\beta^{(M;\alpha'_1,\alpha'_{\tau};m)_{\beta}} d^2\gamma^{(M;\alpha'_1,\alpha'_{\tau};m)_{\gamma}}.
\end{align}
This transformation can be obtained by Eqs.~(\ref{eq:Grassmannintegral}) and (\ref{eq:4Dmodular}).
Namely, the modular symmetry anomalies of Majorana neutrino mass terms are caused by integration of the instanton zero-modes appeared by D-brane instantons.
In the following section, we discuss the detail structure of the modular symmetry anomaly of Majorana mass terms for individual types of models with three generations of neutrinos.

We comment on the automorphy factor $\widetilde{J}_2(\widetilde{\gamma},\tau) $ 
in the anomaly.
As reviewed in section \ref{sec:anomaly}, such anomaly of the automorphy factor can be canceled by 
the Green-Schwarz mechanism due to other moduli $T_\alpha$.
The neutrino mass terms in Eq.~(\ref{eq:neutrinomass}) include the factor $e^{-S_{cl}(T_\alpha,\overline{M}_{inst})}$, 
and this factor does not depend on the complex structure modulus $\tau$ but depends on other moduli $T_\alpha$, 
which correspond to  K\"{a}hler moduli and the dilaton in type IIB string theory.
In the Green-Schwarz mechanism, these moduli $T_\alpha$ transform as Eq.~(\ref{eq:GS})
to cancel the anomaly of the automorphy factor.
Then, it is expected that the modular transformation for $e^{-S_{cl}(T_\alpha,\overline{M}_{inst})}$ 
may cancel the modular weight anomaly, 
i.e. the factor  $\widetilde{J}_2(\widetilde{\gamma},\tau) $ in Eq.~(\ref{eq:n-mass-anomaly}).
In fact, Eq.~(\ref{eq:GS}) implies the shift of the Chern-Simons term in $S_{cl}$, 
and a part of automorphy factor can be cancelled.
At any rate, our purpose is not to show that the 4D Green-Schwarz mechanism works, 
but to show which part of the modular symmetry is anomalous and is broken by non-perturbative neutrino mass terms. 
The modular weight of $m_{a'b'}(\tau) N_{a'}N_{b'}$ without $e^{-S_{cl}}$ 
does not match with other terms in tree-level Lagrangian.
This point is important for 4D modular flavor models.

Furthermore, the factor ${\rm det}\rho_{inst}(\widetilde{\gamma})^{{\bm \alpha}_{inst}{\bm \alpha}'_{inst}}$ 
in Eq.~(\ref{eq:n-mass-anomaly}) can break the flavor symmetry.
As discussed in Ref.~\cite{Kobayashi:2021xfs}, the anomalous part in any non-Abelian discrete group 
corresponds to  $Z_N$ symmetry.
In Ref.~\cite{Kariyazono:2019ehj} it was found that such anomalous $Z_N$ symmetry can be embedded in 
$U(1)$ gauge symmetry.
Note that the neutrino mass terms have $(2,-2)$ charge under $U(1)_1\times U(1)_2$.
That means the D-brane instanton effect breaks $U(1)_1\times U(1)_2$.
As a result, the symmetry $U(1)'=U(1)_1-U(1)_2$ is broken, while the neutrino mass terms are 
invariant under $U''=U(1)_1+U(1)_2$ and this $U(1)''$ symmetry remains.
The Green-Schwarz mechanism requires the moduli $T_\alpha$ in $e^{-S_{cl}(T_\alpha,\overline{M}_{inst})}$
to shift as Eq.(\ref{eq:GSU1}) in order to cancel the  $U(1)'$ anomaly.
When the anomalous part of the modular flavor symmetry 
$\rho_{inst}(\widetilde{\gamma})^{{\bm \alpha}_{inst}{\bm \alpha}'_{inst}}$ corresponds to 
a discrete subgroup $Z_N$ of $U(1)'$ 
as found in Ref.~\cite{Kariyazono:2019ehj}, 
the modular flavor anomaly can be canceled by the same Green-Schwarz mechanism as $U(1)'$.
As said above, our purpose is not to show that the 4D Green-Schwarz mechanism works, 
but to show which part of the modular symmetry is anomalous and is broken by non-perturbative neutrino mass terms. 
Obviously, after the moduli $T_\alpha$ are stabilized, the factor  $e^{-S_{cl}(T_\alpha,\overline{M}_{inst})}$ 
is just a constant.
Then, $U(1)'$ is broken and the subsymmetry of the modular flavor symmetry 
corresponding to ${\rm det}\rho_{inst}(\widetilde{\gamma})^{{\bm \alpha}_{inst}{\bm \alpha}'_{inst}}$ 
in Eq.~(\ref{eq:n-mass-anomaly}) is broken.
Which part is broken in the modular flavor symmetry is important.
We will study it explicitly by use of concrete models in the following section.



\section{Modular flavor symmetry anomalies of Majorana mass terms for three-generation neutrinos}
\label{sec:Modularflavoranomaly}

In this section, we study modular flavor symmetry anomalies of Majorana mass terms for four-types of   models with three generations of neutrinos: the neutrinos with $(M; \alpha_1, \alpha_{\tau}; m)_N=(4;0,0;0)$, $(8;0,0;1)$, $(5;1/2,1/2;1)$, and $(7;1/2,1/2;0)$.
Here, in these models we study non-perturbative neutrino mass terms induced by instanton zero-modes consistent with 
the modular symmetries, i.e. the zero-mode wavefunctions, which transform ones with the same boundary conditions 
under the modular transformation. 
Hereafter, we use the following notations,
\begin{align}
\begin{array}{l}
c^{(M_{\beta},M_{\gamma})} \equiv (2{\rm Im}\tau)^{-1}{\cal A}^{-1} \left( \frac{M_{\beta}M_{\gamma}}{M_{N}} \right)^{1/2},  \\
\eta_{N}^{(M)} \equiv
\vartheta
\begin{bmatrix}
\frac{N}{M} \\ 0
\end{bmatrix}
(0,M\tau), \\
\zeta_{N,L;\pm}^{(M)}
\equiv \eta_{N}^{(M)} \pm_1 \eta_{N+L}^{(M)}, \\
\lambda_{(N,L;\pm_1),K;\pm_2}^{(M)}
\equiv \zeta_{N,L;\pm_1}^{(M)} \pm_2 \zeta_{N+K,L;\pm_1}^{(M)}
\end{array}
\end{align}
and we use the following relation,
\begin{align}
\sum_{k=0}^{g-1} \eta_{gn+(M/g)k}^{(M)} = \eta_{n}^{(M/g^2)},
\end{align}
where $M/g^2 \in \mathbb{Z}$ for $\exists g \in \mathbb{Z}$.


\subsection{Three generations of neutrinos with $(M; \alpha_1, \alpha_{\tau}; m)_N=(4;0,0;0)$}

Here, we study three generations of neutrinos with $(M; \alpha_1, \alpha_{\tau}; m)_N=(4;0,0;0)$.
In this case, the modular transformation matrices for the neutrinos, $\rho_N$ are given as
\begin{align}
\rho_N(\widetilde{S}) = \frac{e^{\pi i/4}}{2}
\begin{pmatrix}
1 & \sqrt{2} & 1 \\
\sqrt{2} & 0 & -\sqrt{2} \\
1 & -\sqrt{2} & 1
\end{pmatrix}, \quad
\rho_N(\widetilde{T}) =
\begin{pmatrix}
1 & 0 & 0 \\
0 & e^{\pi i/4} & 0 \\
0 & 0 & -1
\end{pmatrix},
\end{align}
which satisfy
\begin{align}
\begin{array}{l}
\rho_N(\widetilde{S})^2 = i\mathbb{I}, \\
\rho_N(\widetilde{S})^4 = \left[ \rho_N(\widetilde{S})\rho_N(\widetilde{T}) \right]^3 = -\mathbb{I}, \\
\rho_N(\widetilde{S})^8 = \left[ \rho_N(\widetilde{S})\rho_N(\widetilde{T}) \right]^6 = \rho_N(\widetilde{T})^8 = \mathbb{I},
\end{array}
\label{eq:M4neutrino}
\end{align}
and also 
\begin{align}
\left[ \rho_N(\widetilde{S})^{^1}\rho_N(\widetilde{T})^{-1}\rho_N(\widetilde{S})\rho_N(\widetilde{T}) \right]^3=\mathbb{I}.
\end{align}
They are the unitary representation of $\widetilde{\Delta}(96) \simeq \Delta(48) \rtimes Z_8 \simeq (Z_4 \times Z'_4) \rtimes Z_3 \rtimes Z_8$~\cite{Kikuchi:2021ogn}, where the generators of $Z_4$, $Z'_4$, $Z_3$, and $Z_8$ are given by
\begin{align}
\begin{array}{l}
\rho_N(a) = \rho_N(\widetilde{S}) \rho_N(\widetilde{T})^2 \rho_N(\widetilde{S})^5 \rho_N(\widetilde{T})^4, \\
\rho_N(a') = \rho_N(\widetilde{S}) \rho_N(\widetilde{T})^2 \rho_N(\widetilde{S})^{-1} \rho_N(\widetilde{T})^{-2}, \\
\rho_N(b) = \rho_N(\widetilde{T})^5 \rho_N(\widetilde{S})^5 \rho_N(\widetilde{T})^4, \\
\rho_N(c) = \rho_N(\widetilde{S}) \rho_N(\widetilde{T})^2 \rho_N(\widetilde{S}) \rho_N(\widetilde{T})^5,
\end{array}
\label{eq:delta96}
\end{align}
respectively.

Their Majorana masses can be generated by only one pair of the instanton zero-modes~\cite{Hoshiya:2021nux}, $(\beta, \gamma)^T$, with
\begin{align}
\begin{pmatrix}
(M; \alpha_1, \alpha_{\tau}; m)_{\beta} \\ (M; \alpha_1, \alpha_{\tau}; m)_{\gamma}
\end{pmatrix}
=
\begin{pmatrix}
(2;0,0;0) \\ (2;0,0;0)
\end{pmatrix}. \notag
\end{align}
Here, we denote
$(M,{\bm \alpha})_{inst}=((2,2),(A,A))$. 
Then, the mass matrix can be written as
\begin{align}
M_{ab} &= e^{-S_{cl}(T_\alpha,\overline{M}_{inst}^{(2,2)})} m_{ab}^{((2,2),(A,A))}, \notag \\
&=  e^{-S_{cl}(T_\alpha,\overline{M}_{inst}^{(2,2)})} c^{(2,2)}
\begin{pmatrix}
X_3 & 0 & X_1 \\
0 & -\sqrt{2}X_2 & 0 \\
X_1 & 0 & X_3
\end{pmatrix},
\end{align}
where $X_I\ (I=1,2,3)$ are given by
\begin{align}
\begin{array}{l}
X_1 = (\eta^{(4)}_{0})^2 + (\eta^{(4)}_{2})^2, \qquad 
X_2 = 2\sqrt{2} (\eta^{(4)}_{1})^2 , \qquad 
X_3 = 2 \eta^{(4)}_{0} \eta^{(4)}_{2}.
\end{array}
\end{align}
The modular transformation matrices for the mass matrix elements $(X_1,X_2,X_3)^T \equiv \mathbf{X}^T$, $\rho_m$, are given as
\begin{align}
\rho_m(\widetilde{S}) = \frac{i}{2}
\begin{pmatrix}
1 & \sqrt{2} & 1 \\
\sqrt{2} & 0 & -\sqrt{2} \\
1 & -\sqrt{2} & 1
\end{pmatrix}, \quad
\rho_m(\widetilde{T}) =
\begin{pmatrix}
1 & 0 & 0 \\
0 & i & 0 \\
0 & 0 & -1
\end{pmatrix},
\end{align}
which satisfy 
\begin{align}
\begin{array}{l}
\rho_m(\widetilde{S})^2 = -\mathbb{I}, \\
\rho_m(\widetilde{S})^4 = \left[ \rho_m(\widetilde{S})\rho_m(\widetilde{T}) \right]^3 = \rho_m(\widetilde{T})^4 = \mathbb{I}.
\end{array}
\label{eq:M4mass}
\end{align}
They are the unitary representation of $S'_4 \simeq \Delta'(24) \simeq \Delta(12) \rtimes Z_4 \simeq (Z_2 \times Z'_2) \rtimes Z_3 \rtimes Z_4$~\cite{Hoshiya:2021nux}, where the generators of $Z_2$, $Z'_2$, $Z_3$, and $Z_4$ are the same as Eq.~(\ref{eq:delta96}), respectively, by considering Eq.~(\ref{eq:M4mass}) instead of Eq.~(\ref{eq:M4neutrino}).

On the other hand, since we obtain
\begin{align}
{\rm det}[\rho_{inst}(\widetilde{S})^{(A,A)(A,A)}] = {\rm det}[\rho_{inst}(\widetilde{T})^{(A,A)(A,A)}] = -1,
\end{align}
from Eq.~(\ref{eq:JrhoST}), we can find that
\begin{align}
\begin{array}{l}
{\rm det}[\rho_{inst}(a)^{(A,A)(A,A)}] = {\rm det}[\rho_{inst}(a')^{(A,A)(A,A)}] = {\rm det}[\rho_{inst}(b)^{(A,A)(A,A)}] = 1, \\
{\rm det}[\rho_{inst}(c)^{(A,A)(A,A)}] = -1 \quad ({\rm det}[\rho_{inst}(c^2)^{(A,A)(A,A)}] = 1).
\end{array}
\end{align}
Thus, the Majorana mass term $M_{ab}(\tau) N_a^{(4;0,0;0)}(x) N_b^{(4;0,0;0)}(x)$ is invariant under $a$, $a'$, $b$, and $c^2$ transformation, while it transforms as
\begin{align}
M_{ab}(\tau) N_a^{(4;0,0;0)}(x) N_b^{(4;0,0;0)}(x) \rightarrow - M_{ab}(\tau) N_a^{(4;0,0;0)}(x) N_b^{(4;0,0;0)}(x), \notag
\end{align}
under $c$ transformation.
As a result, among the neutrino flavor symmetry $\widetilde{\Delta}(96) \simeq \Delta(48) \rtimes Z_8$, 
there remains $\Delta(48) \times Z_4$ flavor symmetry in the neutrino mass terms, 
while $Z_2$ part of $Z_8$ symmetry  is broken
\footnote{Actually, according to the analysis in Ref.~\cite{Kobayashi:2021xfs}, we can find that $\Delta(48)$ transformation is automatically anomaly free.}.
Here, the direct product comes from the reason that $\rho_N(c)^2=\rho_N(\widetilde{S})^6=-i\mathbb{I}$~\footnote{See Ref.~\cite{Kikuchi:2021ogn}.} commutes all of the generators, $a$, $a'$, and $b$.


\subsection{Three generations of neutrinos with $(M; \alpha_1, \alpha_{\tau}; m)_N=(8;0,0;1)$}

Here, we study three generations of neutrinos with $(M; \alpha_1, \alpha_{\tau}; m)_N=(8;0,0;1)$.
In this case, the modular transformation matrices for the neutrinos, $\rho_N$ are given as
\begin{align}
\rho_N(\widetilde{S}) = \frac{ie^{\pi i/4}}{2}
\begin{pmatrix}
1 & \sqrt{2} & 1 \\
\sqrt{2} & 0 & -\sqrt{2} \\
1 & -\sqrt{2} & 1
\end{pmatrix}, \quad
\rho_N(\widetilde{T}) = e^{\pi i/8}
\begin{pmatrix}
1 & 0 & 0 \\
0 & e^{3\pi i/8} & 0 \\
0 & 0 & -1
\end{pmatrix},
\end{align}
which satisfy
\begin{align}
\begin{array}{l}
\rho_N(\widetilde{S})^2 = -i\mathbb{I}, \\
\rho_N(\widetilde{S})^4 = \left[ \rho_N(\widetilde{S})\rho_N(\widetilde{T}) \right]^3 = -\mathbb{I}, \\
\rho_N(\widetilde{S})^8 = \left[ \rho_N(\widetilde{S})\rho_N(\widetilde{T}) \right]^6 = \rho_N(\widetilde{T})^{16} = \mathbb{I},
\end{array}
\label{eq:M8neutrino}
\end{align}
and also 
\begin{align}
\left[ \rho_N(\widetilde{S})^{^1}\rho_N(\widetilde{T})^{-1}\rho_N(\widetilde{S})\rho_N(\widetilde{T}) \right]^3=\mathbb{I}.
\end{align}
They are the unitary representation of $\widetilde{\Delta}(384) \simeq \Delta(192) \rtimes Z_8 \simeq (Z_8 \times Z'_8) \rtimes Z_3 \rtimes Z_8$~\cite{Kikuchi:2021ogn}, where the generators of $Z_8$, $Z'_8$, $Z_3$, and $Z_8$ are given by
\begin{align}
\begin{array}{l}
\rho_N(a) = \rho_N(\widetilde{S}) \rho_N(\widetilde{T})^2 \rho_N(\widetilde{S})^5 \rho_N(\widetilde{T})^4, \\
\rho_N(a') = \rho_N(\widetilde{S}) \rho_N(\widetilde{T})^2 \rho_N(\widetilde{S})^{-1} \rho_N(\widetilde{T})^{-2}, \\
\rho_N(b) = \rho_N(\widetilde{T})^7 \rho_N(\widetilde{S})^{11} \rho_N(\widetilde{T})^8, \\
\rho_N(c) = \rho_N(\widetilde{S}) \rho_N(\widetilde{T})^{-10} \rho_N(\widetilde{S}) \rho_N(\widetilde{T})^{-5},
\end{array}
\label{eq:delta384}
\end{align}
respectively.

Note that wavefunctions with $M \in 2\mathbb{Z}$ and $(\alpha_1,\alpha_{\tau})=(0,0)$ and ones with $M \in 2\mathbb{Z}+1$ and $(\alpha_1,\alpha_{\tau})=(1/2,1/2)$ are consistent with the modular symmetry, because they
transform to ones with the same boundary conditions.
The Majorana masses can be generated by two pairs of the instanton zero-modes~\cite{Hoshiya:2021nux}, $(\beta, \gamma)^T$, with
\begin{align}
\begin{pmatrix}
(M; \alpha_1, \alpha_{\tau}; m)_{\beta} \\ (M; \alpha_1, \alpha_{\tau}; m)_{\gamma}
\end{pmatrix}
=&
\begin{pmatrix}
(2;0,0;0) \\ (6;0,0;1)
\end{pmatrix},
\begin{pmatrix}
(3;1/2,1/2;1) \\ (5;1/2,1/2;0) \notag
\end{pmatrix},
\end{align}
which are consistent with the modular transformation.
However, the latter case includes some complexity.
In this section, we study only the former, but we will study the latter in the next section.
Here, we denote
$(M,{\bm \alpha})_{inst}
=
((2,6),(A,A))$.
The mass matrix can be written as
\begin{align}
M_{ab}^{(2,6)}
&= e^{-S_{cl}(T_\alpha,\overline{M}_{inst}^{(2,6)})} m_{ab}^{((2,6),(A,A))} \\
&= e^{-S_{cl}(T_\alpha,\overline{M}_{inst}^{(2,6)})} c^{(2,6)}
\begin{pmatrix}
X_3 & 0 & X_1 \\
0 & -\sqrt{2}X_2 & 0 \\
X_1 & 0 & X_3
\end{pmatrix}, \notag 
\end{align}
where $X_I$ 
 are given by
\begin{align}
&\begin{array}{l}
X_1 = (\zeta_{1,6;-}^{(24)})^2 + (\zeta_{5,6;-}^{(24)})^2, \qquad 
X_2 = \sqrt{2} (\zeta_{2,12;-}^{(24)})^2, \qquad 
X_3 = 2 \zeta_{1,6;-}^{(24)} \zeta_{5,6;-}^{(24)}.
\end{array} 
\end{align}



As discussed in Ref.~\cite{Hoshiya:2021nux}, the modular transformation for the matrix the elements $(X_1,X_2,X_3)^T \equiv \mathbf{X}^T$, $\rho_{m^{(2,6)}}$, is given as
\begin{align}
\rho_{m^{(2,6)}}(\widetilde{S}) = \frac{i}{2}
\begin{pmatrix}
1 & \sqrt{2} & 1 \\
\sqrt{2} & 0 & -\sqrt{2} \\
1 & -\sqrt{2} & 1
\end{pmatrix}, \quad
\rho_{m^{(2,6)}}(\widetilde{T}) = e^{\pi i/12}
\begin{pmatrix}
1 & 0 & 0 \\
0 & e^{\pi i/4} & 0 \\
0 & 0 & -1
\end{pmatrix},
\end{align}
and they satisfy
\begin{align}
\begin{array}{l}
\rho_{m^{(2,6)}}(\widetilde{S})^2 = -\mathbb{I}, \quad \rho_{m^{(2,6)}}(\widetilde{T})^8 = e^{2\pi i/3} \mathbb{I}, \\
\rho_{m^{(2,6)}}(\widetilde{S})^4 = \left[ \rho_{m^{(2,6)}}(\widetilde{S})\rho_{m^{(2,6)}}(\widetilde{T}) \right]^3 =  \rho_{m^{(2,6)}}(\widetilde{T})^{24} = \mathbb{I},
\end{array}
\label{eq:M826mass}
\end{align}
and also satisfy
\begin{align}
\left[ \rho_{m^{(2,6)}}(\widetilde{S})^{-1}\rho_{m^{(2,6)}}(\widetilde{T})^{-1}\rho_{m^{(2,6)}}(\widetilde{S})\rho_{m^{(2,6)}}(\widetilde{T}) \right]^3=\mathbb{I}.
\end{align}
They are the unitary representation of $\Delta'(96) \times Z_3 \simeq (\Delta(48) \rtimes Z_4) \times Z_3 \simeq ((Z_4 \times Z'_4) \rtimes Z_3 \rtimes Z_4 ) \times Z_3$, where the generators of $Z_4$, $Z'_4$, $Z_3$, and $Z_4$ are the same as  Eq.~(\ref{eq:delta384}), respectively, by considering Eq.~(\ref{eq:M826mass}) instead of Eq.~(\ref{eq:M8neutrino}), while the generator of the last $Z_3$ is given by
\begin{align}
\rho_{m^{(2,6)}}(d) = \rho_{m^{(2,6)}}(\widetilde{T})^{16}.
\end{align}
Note that this transformation for the neutrinos is trivial: $\rho_N(d)=\mathbb{I}$.

On the other hand, since we obtain
\begin{align}
{\rm det}[\rho_{inst^{(2,6)}}(\widetilde{S})^{(A,A)(A,A)}] = 1, \quad {\rm det}[\rho_{inst^{(2,6)}}(\widetilde{T})^{(A,A)(A,A)}] = e^{4\pi i/3},
\end{align}
from Eq.~(\ref{eq:JrhoST}), we can find that
\begin{align}
\begin{array}{l}
{\rm det}[\rho_{inst^{(2,6)}}(a)^{(A,A)(A,A)}] = {\rm det}[\rho_{inst^{(2,6)}}(a')^{(A,A)(A,A)}] \\
= {\rm det}[\rho_{inst^{(2,6)}}(b)^{(A,A)(A,A)}] = {\rm det}[\rho_{inst^{(2,6)}}(c)^{(A,A)(A,A)}] = 1, \\
{\rm det}[\rho_{inst^{(2,6)}}(d)^{(A,A)(A,A)}] = e^{4\pi i/3} \quad ({\rm det}[\rho_{inst^{(2,6)}}(d^3)^{(A,A)(A,A)}] = 1).
\label{eq:inst26}
\end{array}
\end{align}
Thus, the Majorana mass term generated by instanton zero-modes with $(M_{\beta}, M_{\gamma})=(2,6)$, $M_{ab}^{(2,6)}(\tau) N_a^{(8;0,0;1)}(x) N_b^{(8;0,0;1)}(x)$ is invariant under $a$, $a'$, $b$, $c$, and $d^3$ transformation, while it transforms as
\begin{align}
M_{ab}^{(2,6)}(\tau) N_a^{(8;0,0;1)}(x) N_b^{(8;0,0;1)}(x) \rightarrow e^{4\pi i/3} M_{ab}^{(2,6)}(\tau) N_a^{(8;0,0;1)}(x) N_b^{(8;0,0;1)}(x), \notag
\end{align}
under $d$ transformation.
As a result, the full $\widetilde{\Delta}(384)$ flavor symmetry of neutrinos remains, although $Z_3$ transformation for $M_{ab}^{(2,6)}(\tau)$ becomes meaningless in Lagrangian.



\subsection{Three generations of neutrinos with $(M; \alpha_1, \alpha_{\tau}; m)_N=(5;1/2,1/2;1)$}

Here, we study three generations of neutrinos with $(M; \alpha_1, \alpha_{\tau}; m)_N=(5;1/2,1/2;1)$.
In this case, the modular transformation matrices for the neutrinos, $\rho_N$ are given as
\begin{align}
\rho_N(\widetilde{S}) = \frac{ie^{\pi i/4}}{\sqrt{5}}
\begin{pmatrix}
2s(1) & 2s(3) & \sqrt{2} \\
2s(3) & 2s(1) & -\sqrt{2} \\
\sqrt{2} & -\sqrt{2} & 1
\end{pmatrix}, \quad
\rho_N(\widetilde{T}) =
\begin{pmatrix}
e^{\pi i/20} & 0 & 0 \\
0 & e^{9\pi i/20} & 0 \\
0 & 0 & e^{25\pi i/20}
\end{pmatrix},
\end{align}
and they satisfy
\begin{align}
\begin{array}{l}
\rho_N(\widetilde{T})^5 = e^{\pi i/4} \mathbb{I}, \\
\rho_N(\widetilde{S})^2 = -i\mathbb{I}, \quad \rho_N(\widetilde{T})^{10} = i\mathbb{I}, \\
\rho_N(\widetilde{S})^4 = \left[ \rho_N(\widetilde{S})\rho_N(\widetilde{T}) \right]^3 = \rho_N(\widetilde{T})^{20} =  -\mathbb{I}, \\
\rho_N(\widetilde{S})^8 = \left[ \rho_N(\widetilde{S})\rho_N(\widetilde{T}) \right]^6 = \rho_N(\widetilde{T})^{40} = \mathbb{I},
\end{array}
\label{eq:M5neutrino}
\end{align}
where $s(n) \equiv \sin (n\pi/10)$.
They are the unitary representation of $A_5 \times Z_8$~\cite{Kikuchi:2021ogn}, where the generators of $A_5$ are given by
\begin{align}
\begin{array}{l}
\rho_N(S') = \rho_N(\widetilde{S}) \rho_N(\widetilde{T})^{45}, \qquad 
\rho_N(T') = \rho_N(\widetilde{T})^{-24}, \\
\end{array}
\label{eq:A5}
\end{align}
and the generator of $Z_8$ is given by
\begin{align}
\rho_N(c) = \rho_N(T)^5.
\label{eq:A5centerZ8}
\end{align}

Their Majorana masses can be generated by only one pair of the instanton zero-modes, $(\beta, \gamma)^T$, with
\begin{align}
\begin{pmatrix}
(M; \alpha_1, \alpha_{\tau}; m)_{\beta} \\ (M; \alpha_1, \alpha_{\tau}; m)_{\gamma}
\end{pmatrix}
=
\begin{pmatrix}
(2;0,0;0) \\ (3;1/2,1/2;1)
\end{pmatrix}. \notag
\end{align}
Here, we denote
$(M,{\bm \alpha})_{inst}=((2,3),(A,D))$. 
Then, the mass matrix can be written as
\begin{align}
M_{ab} &= e^{-S_{cl}(T_\alpha,\overline{M}_{inst}^{(2,3)})} m_{ab}^{((2,3),(A,D))}, \notag \\
&= e^{-S_{cl}(T_\alpha,\overline{M}_{inst}^{(2,3)})} c^{2,3}
\begin{pmatrix}
\sqrt{2} X_1 & X_4 & X_5 \\
X_4 & \sqrt{2} X_2 & X_6 \\
X_5 & X_6 & \sqrt{2} X_3
\end{pmatrix},
\end{align}
where $X_I\ (I=1,2,3,4,5,6)$ are given by
\begin{align}
\begin{array}{l}
X_1 = 2 ( \zeta_{1,10;+}^{(30)} \eta_{6}^{(30)}  - \zeta_{4,10;+}^{(30)} \eta^{(30)}_9 ), \\
X_2 = 2 ( \zeta_{7,10;+}^{(30)} \eta_{12}^{(30)}  - \zeta_{-2,10;+}^{(30)} \eta^{(30)}_{3} ), \\
X_3 = 2 (\eta_{0}^{(30)}  \eta_{5}^{(30)}  -  \eta^{(30)}_{15} \eta^{(30)}_{10}), \\
X_4 = \sqrt{2} ( \zeta_{-2,10;+}^{(30)} \eta_{9}^{(30)}  -  \zeta_{7,10;+}^{(30)} \eta^{(30)}_{6} + \zeta_{4,10;+}^{(30)} \eta^{(30)}_{3} - \zeta_{1,10;+}^{(30)} \eta^{(30)}_{12} ), \\
X_5 = ( \zeta_{1,10;+}^{(30)} \eta^{(30)}_{0} - \zeta_{4,10;+}^{(30)} \eta^{(30)}_{15} + 2 \eta_{5}^{(30)} \eta_{6}^{(30)} - 2 \eta^{(30)}_{10} \eta^{(30)}_{9} ), \\
X_6 = ( \zeta_{-2,10;+}^{(30)} \eta^{(30)}_{15} - \zeta_{7,10;+}^{(30)} \eta^{(30)}_{0} + 2 \eta_{3}^{(30)} \eta_{10}^{(30)} - 2 \eta^{(30)}_{12} \eta^{(30)}_{5} ).
\end{array}
\end{align}
The modular transformation matrices for the mass matrix elements $(X_1,X_2,X_3,X_4,X_5,X_6)^T \equiv \mathbf{X}^T$, $\rho_m$, are given as
\begin{align}
\begin{array}{l}
\rho_m(\widetilde{S}) = \frac{i}{5}
\begin{pmatrix}
4s^2(1) & 4s^2(3) & 2 & \sqrt{2} & 4s(1) & 4s(3) \\
4s^2(3) & 4s^2(1) & 2 & \sqrt{2} & -4s(1) & -4s(3) \\
2 & 2 & 1 & -2\sqrt{2} & 2 & -2 \\
\sqrt{2} & \sqrt{2} & -2\sqrt{2} & 3 & \sqrt{2} & -\sqrt{2} \\
4s(1) & -4s(3) & 2 & \sqrt{2} & 2+2s(1) & 2+2s(3) \\
4s(3) & -4s(1) & -2 & \sqrt{2} & -2+2s(3) & 2+2s(1)
\end{pmatrix}, \\
\rho_m(\widetilde{T}) =
\begin{pmatrix}
e^{-23 \pi i/30} & 0 & 0 & 0 & 0 & 0 \\
0 & e^{13 \pi i/30} & 0 & 0 & 0 & 0 \\
0 & 0 &  e^{25 \pi i/30} & 0 & 0 & 0 \\
0 & 0 & 0 &  e^{25 \pi i/30} & 0 & 0 \\
0 & 0 & 0 & 0 &  e^{\pi i/30} & 0 \\
0 & 0 & 0 & 0 & 0 &  e^{-11 \pi i/30}
\end{pmatrix},
\end{array}
\end{align}
which satisfy
\begin{align}
\begin{array}{l}
\rho_m(\widetilde{T})^5 = e^{\pi i/6} \mathbb{I}, \\
\rho_m(\widetilde{S})^2 = \rho_m(\widetilde{T})^{30} = -\mathbb{I}, \\
\rho_m(\widetilde{S})^4 = \left[ \rho_m(\widetilde{S})\rho_m(\widetilde{T}) \right]^3 = \rho_m(\widetilde{T})^{60} = \mathbb{I}.
\end{array}
\label{eq:M5mass}
\end{align}
They are the unitary representation of $A_5 \times Z_{12}$, where the generators of $A_5$ and $Z_{12}$ are the same as  Eqs.~(\ref{eq:A5}) and (\ref{eq:A5centerZ8}), respectively, by considering Eq.~(\ref{eq:M5mass}) instead of Eq.~(\ref{eq:M5neutrino}).

On the other hand, since we obtain
\begin{align}
{\rm det}[\rho_{inst}(\widetilde{S})^{(A,D)(A,D)}] = 1, \quad {\rm det}[\rho_{inst}(\widetilde{T})^{(A,D)(A,D)}] = e^{4\pi i/3},
\end{align}
from Eq.~(\ref{eq:JrhoST}), we can find that
\begin{align}
\begin{array}{l}
{\rm det}[\rho_{inst}(S')^{(A,D)(A,D)}] = {\rm det}[\rho_{inst}(T')^{(A,D)(A,D)}] = 1, \\
{\rm det}[\rho_{inst}(c)^{(A,D)(A,D)}] = e^{2\pi i/3} \quad ({\rm det}[\rho_{inst}(c^3)^{(A,D)(A,D)}] = 1).
\end{array}
\end{align}
Thus, the Majorana mass term $M_{ab}(\tau) N_a^{(5;1/2,1/2;1)}(x) N_b^{(5;1/2,1/2;1)}(x)$ is invariant under $S'$, $T'$, and $c^3$ transformation, while it transforms as
\begin{align}
M_{ab}(\tau) N_a^{(5;1/2,1/2;1)}(x) N_b^{(5;1/2,1/2;1)}(x) \rightarrow e^{2\pi i/3} M_{ab}(\tau) N_a^{(5;1/2,1/2;1)}(x) N_b^{(5;1/2,1/2;1)}(x), \notag
\end{align}
under $c$ transformation.
Note that the anomaly free $c^3$ transformation becomes the generator of $Z_8$ symmetry for neutrinos.
As a result, the full $A_5 \times Z_8$ flavor symmetry of neutrinos  remain, although $Z_3$ transformation of $Z_{12}$  for the mass matrix elements becomes meaningless in Lagrangian\footnote{Actually, according to the analysis in Ref.~\cite{Kobayashi:2021xfs}, we can find that $A_5$ transformation is automatically anomaly free.}.


\subsection{Three generations of neutrinos with $(M; \alpha_1, \alpha_{\tau}; m)_N=(7;1/2,1/2;0)$}

Here, we study three generations of neutrinos with $(M; \alpha_1, \alpha_{\tau}; m)_N=(7;1/2,1/2;0)$.
In this case, the modular transformation matrix for the neutrinos, $\rho_N$ is given as
\begin{align}
\rho_N(\widetilde{S}) = \frac{2e^{\pi i/4}}{\sqrt{7}}
\begin{pmatrix}
c(1) & c(3) & c(5) \\
c(3) & c(9) & -c(1) \\
c(5) & -c(1) & c(3)
\end{pmatrix}, \quad
\rho_N(\widetilde{T}) = 
\begin{pmatrix}
e^{\pi i/28} & 0 & 0 \\
0 & e^{9\pi i/28} & 0 \\
0 & 0 & e^{25\pi i/28}
\end{pmatrix},
\end{align}
where $c(n) \equiv \cos (n\pi/14)$.
They satisfy
\begin{align}
\begin{array}{l}
\rho_N(\widetilde{T})^7 = e^{\pi i/4} \mathbb{I}, \\
\rho_N(\widetilde{S})^2 = \rho_N(\widetilde{T})^{14} = i\mathbb{I}, \\
\rho_N(\widetilde{S})^4 = \left[ \rho_N(\widetilde{S})\rho_N(\widetilde{T}) \right]^3 = \rho_N(\widetilde{T})^{28} =  -\mathbb{I}, \\
\rho_N(\widetilde{S})^8 = \left[ \rho_N(\widetilde{S})\rho_N(\widetilde{T}) \right]^6 = \rho_N(\widetilde{T})^{56} = \mathbb{I},
\end{array}
\label{eq:M7neutrino}
\end{align}
and also satisfy
\begin{align}
\left[ \rho_N(\widetilde{S})^{^1}\rho_N(\widetilde{T})^{-1}\rho_N(\widetilde{S})\rho_N(\widetilde{T}) \right]^4=\mathbb{I}.
\end{align}
They are  the unitary representation of $PSL(2,Z_7) \times Z_8$~\cite{Kikuchi:2021ogn}, where the generators of $PSL(2,Z_7)$ are given by
\begin{align}
\begin{array}{l}
\rho_N(S') = \rho_N(\widetilde{S}) \rho_N(\widetilde{T})^{21}, \\
\rho_N(T') = \rho_N(\widetilde{T})^{24}, \\
\end{array}
\label{eq:PSL2Z7}
\end{align}
and the generator of $Z_8$ is given by
\begin{align}
\rho_N(c) = \rho_N(T)^7.
\label{eq:PSL2Z7centerZ8}
\end{align}

Their Majorana masses can be generated by one pair of the instanton zero-modes, $(\beta, \gamma)^T$, with
\begin{align}
\begin{pmatrix}
(M; \alpha_1, \alpha_{\tau}; m)_{\beta} \\ (M; \alpha_1, \alpha_{\tau}; m)_{\gamma}
\end{pmatrix}
=
\begin{pmatrix}
(2;0,0;0) \\ (5;1/2,1/2;0) \notag
\end{pmatrix},
\end{align}
which is consistent with the modular transformation.
Then, the mass matrix can be written as
\begin{align}
M_{ab}^{(2,5)}
&= e^{-S_{cl}(T_\alpha,\overline{M}_{inst}^{(2,5)})} m_{ab}^{((2,5),(A,D))} \\
&= e^{-S_{cl}(T_\alpha,\overline{M}_{inst}^{(2,5)})} c^{(2,5)}
\begin{pmatrix}
\sqrt{2} X_1 & X_4 & X_5 \\
X_4 & \sqrt{2}X_2 & X_6 \\
X_5 & X_6 & \sqrt{2} X_3
\end{pmatrix}, \notag \\
\end{align}
where $X_I$ are given by
\begin{align}
&\begin{array}{l}
X_1 = \sqrt{2} ( \zeta_{-1,30;-}^{(70)} \zeta_{-8,30;-}^{(70)} - \zeta_{13,30;-}^{(70)} \zeta_{6,30;-}^{(70)} ), \\
X_2 = \sqrt{2} ( \zeta_{11,20;-}^{(70)} \zeta_{18,20;-}^{(70)} - \zeta_{-3,20;-}^{(70)} \zeta_{4,20;-}^{(70)} ), \\
X_3 = \sqrt{2} ( \zeta_{9,10;-}^{(70)} \zeta_{2,10;-}^{(70)} - \zeta_{23,10;-}^{(70)} \zeta_{16,10;-}^{(70)} ), \\
X_4 = ( \zeta_{-1,30;-}^{(70)} \zeta_{18,20;-}^{(70)} + \zeta_{-8,30;-}^{(70)} \zeta_{11,20;-}^{(70)} - \zeta_{13,30;-}^{(70)} \zeta_{4,20;-}^{(70)} - \zeta_{6,30;-}^{(70)} \zeta_{-3,20;-}^{(70)} ), \\
X_5 = ( \zeta_{-1,30;-}^{(70)} \zeta_{2,10;-}^{(70)} + \zeta_{-8,30;-}^{(70)} \zeta_{9,10;-}^{(70)} - \zeta_{13,30;-}^{(70)} \zeta_{16,10;-}^{(70)} - \zeta_{6,30;-}^{(70)} \zeta_{23,10;-}^{(70)} ), \\
X_6 = ( \zeta_{11,20;-}^{(70)} \zeta_{2,10;-}^{(70)} + \zeta_{18,20;-}^{(70)} \zeta_{9,10;-}^{(70)} - \zeta_{-3,20;-}^{(70)} \zeta_{16,10;-}^{(70)} - \zeta_{4,20;-}^{(70)} \zeta_{23,10;-}^{(70)} ).
\end{array}
\end{align}
The modular transformation matrix for the mass matrix elements $(X_1,X_2,X_3,X_4,X_5,X_6)^T \equiv \mathbf{X}^T$, $\rho_{m^{(2,5)}}$, is given as
\begin{align}
\begin{array}{l}
\rho_{m^{(2,5)}}(\widetilde{S}) = \frac{4i}{7}
\begin{pmatrix}
S_1 & S_2 \\
S_2^T & S_3
\end{pmatrix}, \\
\rho_{m^{(2,5)}}(\widetilde{T}) =
\begin{pmatrix}
e^{13 \pi i/14} & 0 & 0 & 0 & 0 & 0 \\
0 & e^{5 \pi i/14} & 0 & 0 & 0 & 0 \\
0 & 0 & e^{17 \pi i/14} & 0 & 0 & 0 \\
0 & 0 & 0 & e^{9 \pi i/14} & 0 & 0 \\
0 & 0 & 0 & 0 & e^{\pi i/14} & 0 \\
0 & 0 & 0 & 0 & 0 & e^{25 \pi i/14}
\end{pmatrix},
\end{array}
\end{align}
where $S_i\ (i=1,2,3)$ denote
\begin{align}
\begin{array}{l}
S_1 =
\begin{pmatrix}
c^2(1) & c^2(3) & c^2(5) \\
c^2(3) & c^2(5) & c^2(1) \\
c^2(1) & c^2(1) & c^2(3)
\end{pmatrix}, \\
S_2 =
\begin{pmatrix}
\sqrt{2}c(1)c(3) & \sqrt{2}c(1)c(5) & \sqrt{2}c(3)c(5) \\
-\sqrt{2}c(3)c(5) & -\sqrt{2}c(1)c(3) & \sqrt{2}c(1)c(5) \\
-\sqrt{2}c(1)c(5) & \sqrt{2}c(3)c(5) & -\sqrt{2}c(1)s(3)
\end{pmatrix}, \\
S_3 =
\begin{pmatrix}
c^2(3)-c(1)c(5) & -c^2(1)+c(3)c(5) & -c^2(5)-c(1)c(3) \\
-c^2(1)+c(3)c(5) & c^2(5)+c(1)c(3) & c^2(3)-c(1)c(5) \\
-c^2(5)-c(1)c(3) & c^2(3)-c(1)c(5) & c^2(1)-c(3)c(5) 
\end{pmatrix}.
\end{array}
\end{align}
They satisfy
\begin{align}
\begin{array}{l}
\rho_{m^{(2,5)}}(\widetilde{T})^7 = i \mathbb{I}, \\
\rho_{m^{(2,5)}}(\widetilde{S})^2 = \rho_{m^{(2,5)}}(\widetilde{T})^{14} = -\mathbb{I}, \\
\rho_{m^{(2,5)}}(\widetilde{S})^4 = \left[ \rho_{m^{(2,5)}}(\widetilde{S})\rho_{m^{(2,5)}}(\widetilde{T}) \right]^3 = \rho_{m^{(2,5)}}(\widetilde{T})^{28} = \mathbb{I}.
\end{array}
\label{eq:M725mass}
\end{align}
They are the unitary representation of $PSL(2,Z_7) \times Z_{4}$, where the generators of $PSL(2,Z_7)$ and $Z_{4}$ are the same as  Eqs.~(\ref{eq:PSL2Z7}) and (\ref{eq:PSL2Z7centerZ8}), respectively, by considering Eq.~(\ref{eq:M725mass}) instead of Eq.~(\ref{eq:M7neutrino}).

On the other hand, since we obtain
\begin{align}
{\rm det}[\rho_{inst^{(2,5)}}(\widetilde{S})^{(A,D)(A,D)}] = {\rm det}[\rho_{inst^{(2,5)}}(\widetilde{T})^{(A,D)(A,D)}] = -1,
\end{align}
from Eq.~(\ref{eq:JrhoST}), we can find that
\begin{align}
\begin{array}{l}
{\rm det}[\rho_{inst^{(2,5)}}(S')^{(A,D)(A,D)}] = {\rm det}[\rho_{inst^{(2,5)}}(T')^{(A,D)(A,D)}] = 1, \\
{\rm det}[\rho_{inst^{(2,5)}}(c)^{(A,D)(A,D)}] = -1 \quad ({\rm det}[\rho_{inst^{(2,5)}}(c^2)^{(A,D)(A,D)}] = 1).
\end{array}
\end{align}
Thus, the Majorana mass term $M_{ab}^{(2,5)}(\tau) N_a^{(7;1/2,1/2;0)}(x) N_b^{(7;1/2,1/2;0)}(x)$ is invariant under $S'$, $T'$, and $c^2$ transformation, while it transforms as
\begin{align}
M_{ab}^{(2,5)}(\tau) N_a^{(7;1/2,1/2;0)}(x) N_b^{(7;1/2,1/2;0)}(x) \rightarrow - M_{ab}^{(2,5)}(\tau) N_a^{(7;1/2,1/2;0)}(x) N_b^{(7;1/2,1/2;0)}(x), \notag
\end{align}
under $c$ transformation.
As a result, among the neutrino flavor symmetry $PSL(2,Z_7) \times Z_8$, 
there remains $PSL(2,Z_7) \times Z_4$ flavor symmetry in neutrino mass terms, 
while $Z_2$ part of $Z_{8}$ symmetry is broken \footnote{Actually, according to the analysis in Ref.~\cite{Kobayashi:2021xfs}, we can find that $PSL(2,Z_7)$ transformation is automatically anomaly free.}.


\subsection{Results and implications to phenomenology}

Here, we examine our results and implications to particle phenomenology.
First, the modular weight of neutrino masses terms does not match with other terms in tree-level Lagrangian.
We consider the following superpotential in supersymmetric model,
\begin{align}
W=AY_{ab}(\tau)L_aH_uN_b + Bm_{ab}(\tau)N_aN_b,
\end{align}
where $L_a, N_n, H_u$ denote superfields of left-handed leptons, right-handed neutrinos, and Higgs field, 
$Y_{ab}(\tau)$ and $m_{ab}(\tau)$ are modular forms corresponding to Yukawa couplings and neutrino masses, 
respectively.
Here, $A$ and $B$ are just constants, which are written following the convention of 
recent 4D modular flavor models \cite{Feruglio:2017spp}, although $A$ and $B$ may depend on other moduli $T_\alpha$ 
in string-derived low energy effective field theory.
Supoose that the first Yukawa terms are tree-level terms, while 
the second neutrino masses terms induced by non-perturbative effects.
In global supersymmetric models, we require the modular invariance of tree-level terms.
That is, the Yukawa terms $Y_{ab}(\tau)L_aH_dN_b$ have totally vanishing modular weight.
However, our analysis shows that the neutrino mass terms $m_{ab}(\tau)N_aN_b$ have 
non-vanishing modular weight.
In our $T^2/Z_2$ orbifold models, its modular weight is two.
In general, the modular weight of he neutrino mass terms $m_{ab}(\tau)N_aN_b$ would depend on 
compactification, e.g. the sum of modular weights of zero-modes, $\beta_i$ and $\gamma_i$ ,
which is $(-2)$ times the modular weight of $N_aN_b$.

Next, the tree-level flavor symmetry can break to its normal subgroup in the neutrino mass terms, 
although there is the example, where the full tree-level flavor symmetry remains.
Suppose that there is the flavor symmetry $G\rtimes Z_N$ at tree level.
Non-perturbative effects may break $Z_N$, and only the flavor symmetry $G$ may remain 
in the neutrino mass terms.
For example, suppose that the tree level flavor symmetry is $S_4$.
It may break to $A_4$ in neutrino mass terms\footnote{
A similar scenario was studied in Refs.~\cite{Kobayashi:2019mna,Kobayashi:2019xvz}, although such breaking effects were included in the Yukawa sector .}.


\section{More corrections}
\label{sec:more}

In the previous section, we have studied the neutrino mass terms induced by 
the D-brane instanton whose zero-modes transform the ones with the same boundary conditions 
under the modular symmetry.
Note that wavefunctions with $M \in 2\mathbb{Z}$ and $(\alpha_1,\alpha_{\tau})=(0,0)$ and ones with $M \in 2\mathbb{Z}+1$ and $(\alpha_1,\alpha_{\tau})=(1/2,1/2)$ are consistent with the modular symmetry, because they
transform to ones with the same boundary conditions.
Wavefunctions with other SS phases transform to ones with different SS phases.
Such D-brane instanton zero-modes are allowed by requiring only the condition (\ref{eq:zero-mode-condition}),
although consistency with the modular symmetry may forbid such zero-modes.
Here, we attempt to investigate contributions due to such zero-modes.

For the neutrinos with $(M; \alpha_1, \alpha_{\tau}; m)_N=(4;0,0;0)$, and $(5;1/2,1/2;1)$, 
all the possible D-brane instanton zero-modes satisfying the condition (\ref{eq:zero-mode-condition}) 
have been studied in the previous section, 
and all of them are consistent with the modular transformation.
On the other hand, the neutrinos with $(M; \alpha_1, \alpha_{\tau}; m)_N=(8;0,0;1)$, and $(7;1/2,1/2;0)$ 
have other possibilities for D-brane instanton zero-modes satisfying the condition (\ref{eq:zero-mode-condition}).
Here, we study them.



\subsection{Three generations of neutrinos with $(M; \alpha_1, \alpha_{\tau}; m)_N=(8;0,0;1)$}

The possible instanton zero-modes satisfying the condition (\ref{eq:zero-mode-condition}) 
are $(\beta, \gamma)^T$ with~\cite{Hoshiya:2021nux}, 
\begin{align}
\begin{pmatrix}
(M; \alpha_1, \alpha_{\tau}; m)_{\beta} \\ (M; \alpha_1, \alpha_{\tau}; m)_{\gamma}
\end{pmatrix}
=&
\begin{pmatrix}
(2;0,0;0) \\ (6;0,0;1)
\end{pmatrix},
\begin{pmatrix}
(3;0,0;0) \\ (5;0,0;1)
\end{pmatrix},
\begin{pmatrix}
(3;1/2,0;0) \\ (5;1/2,0;1)
\end{pmatrix},
\begin{pmatrix}
(3;0,1/2;0) \\ (5;0,1/2;1)
\end{pmatrix}, \notag \\&
\begin{pmatrix}
(3;1/2,1/2;1) \\ (5;1/2,1/2;0)
\end{pmatrix},
\begin{pmatrix}
(4;1/2,0;0) \\ (4;1/2,0;1)
\end{pmatrix},
\begin{pmatrix}
(4;0,1/2;0) \\ (4;0,1/2;1)
\end{pmatrix},
\begin{pmatrix}
(4;1/2,1/2;0) \\ (4;1/2,1/2;1)
\end{pmatrix}. \notag
\end{align}
Here, we denote
\begin{align}
(M,{\bm \alpha})_{inst}
=&
((2,6),(A,A)), ((3,5),(A,A)), ((3,5),(B,B)), ((3,5),(C,C)), \notag \\&
((3,5),(D,D)), ((4,4),(B,B)), ((4,4),(C,C)), ((4,4),(D,D)), \notag
\end{align}
respectively.
Then, the matrix can be written as
\begin{align}
M_{ab}
&= M_{ab}^{(2,6)} + M_{ab}^{(3,5)} + M_{ab}^{(4,4)}, \label{eq:M8total} \\
M_{ab}^{(3,5)} 
&= e^{-S_{cl}(T_\alpha,\overline{M}_{inst}^{(3,5)})} \sum_{{\bm \alpha}_{inst}=(A,A),(B,B),(C,C),(D,D)} m_{ab}^{((3,5),{\bm \alpha}_{inst})}, \\
&= e^{-S_{cl}(T_\alpha,\overline{M}_{inst}^{(3,5)})} c^{(3,5)}
\begin{pmatrix}
Y_3 & 0 & Y_1 \\
0 & -\sqrt{2}Y_2 & 0 \\
Y_1 & 0 & Y_3
\end{pmatrix}, \notag \\
M_{ab}^{(4,4)}
&= e^{-S_{cl}(T_\alpha,\overline{M}_{inst}^{(4,4)})} \sum_{{\bm \alpha}_{inst}=(B,B),(C,C),(D,D)} m_{ab}^{((4,4),{\bm \alpha}_{inst})} \\
&= e^{-S_{cl}(T_\alpha,\overline{M}_{inst}^{(4,4)})} c^{(4,4)}
\begin{pmatrix}
Z_3 & 0 & Z_1 \\
0 & - \sqrt{2} Z_2 & 0 \\
Z_1 & 0 & Z_3
\end{pmatrix}, \notag
\end{align}
where $Y_I$, $Z_I \equiv Z'_I + Z''_I$ $(I=1,2,3)$ are given by
\begin{align}
&\begin{array}{l}
Y_1 = 2\sqrt{2} ( \zeta_{9,30;-}^{(120)} \lambda_{(37,30;-),40;+}^{(120)} + \zeta_{33,30;-}^{(120)} \lambda_{(-11,30;-),40;+}^{(120)} + \zeta_{-3,30;-}^{(120)} \lambda_{(1,30;-),40;+}^{(120)} + \zeta_{21,30;-}^{(120)} \lambda_{(-47,30;-),40;+}^{(120)} ) \\
Y_2 = 4 ( \zeta^{(120)}_{-6,60;-} \lambda^{(120)}_{(-2,60;-),40;+} - \zeta^{(120)}_{-18,60;-} \lambda^{(120)}_{(-14,60;-),40;+} ), \\
Y_3 = 2\sqrt{2} ( \zeta_{9,30;-}^{(120)} \lambda_{(-47,30;-),40;+}^{(120)} + \zeta_{33,30;-}^{(120)} \lambda_{(1,30;-),40;+}^{(120)} + \zeta_{21,30;-}^{(120)} \lambda_{(37,30;-),40;+}^{(120)} + \zeta_{-3,30;-}^{(120)} \lambda_{(-11,30;-),40;+}^{(120)} )
\end{array} \\
&\begin{array}{ll}
Z'_1 = (\eta^{(8)}_{1})^2 + (\eta^{(8)}_{3})^2, & Z''_1 = (\eta^{(8)}_{0})^2 + (\eta^{(8)}_{2})^2, \\
Z'_2 = \sqrt{2} \eta^{(8)}_{2} (\eta^{(8)}_{0} + \eta^{(8)}_{4}), & Z''_2 = \sqrt{2} \left( (\eta^{(8)}_{1})^2 + (\eta^{(8)}_{3})^2 \right), \\
Z'_3 = 2 \eta^{(8)}_{1} \eta^{(8)}_{3}, & Z''_3 = 2 (\eta^{(8)}_{2})^2,
\end{array}
\end{align}
respectively.
Note that $M_{ab}^{(2,6)}$ is already obtained in the previous section.



\subsubsection{$(M_{\beta},M_{\gamma})=(3,5)$ case}

Let us study the case of the instanton zero-modes with $(M_{\beta},M_{\gamma})=(3,5)$.
In this case, the modular transformation for the matrix elements $(Y_1,Y_2,Y_3)^T \equiv \mathbf{Y}^T$, $\rho_{m^{(3,5)}}$, is the same as $\rho_{m^{(2,6)}}$,~i.e. $\rho_{m^{(3,5)}} = \rho_{m^{(2,6)}}$.
Then, this also becomes the unitary representation of $\Delta'(96) \times Z_3$.

On the other hand, we obtain
\begin{align}
\begin{array}{l}
{\rm det}[\rho_{inst^{(3,5)}}(\widetilde{S})^{(A,A)(A,A)}] = {\rm det}[\rho_{inst^{(3,5)}}(\widetilde{S})^{(B,B)(C,C)}] \\
= {\rm det}[\rho_{inst^{(3,5)}}(\widetilde{S})^{(C,C)(B,B)}] = {\rm det}[\rho_{inst^{(3,5)}}(\widetilde{S})^{(D,D)(D,D)}] = 1, \\
{\rm det}[\rho_{inst^{(3,5)}}(\widetilde{T})^{(A,A)(C,C)}] = {\rm det}[\rho_{inst^{(3,5)}}(\widetilde{T})^{(B,B)(B,B)}] \\
= {\rm det}[\rho_{inst^{(3,5)}}(\widetilde{T})^{(C,C)(A,A)}] = {\rm det}[\rho_{inst^{(3,5)}}(\widetilde{T})^{(D,D)(D,D)}] = e^{4\pi i/3},
\end{array}
\end{align}
from Eq.~(\ref{eq:JrhoST}), and then we find
\begin{align}
{\rm det}[\rho_{inst^{(3,5)}}(\widetilde{S})^{(total)(total)}] = 1, \quad {\rm det}[\rho_{inst^{(3,5)}}(\widetilde{T})^{(total)(total)}] = e^{4\pi i/3},
\end{align}
which are the same as Eq.~(\ref{eq:inst26}).
Note that this is the reason why the elements of $M_{ab}^{(3,5)}$, $\mathbf{Y}^T$, are closed under the modular transformation.
Thus, the results is the same as 
the case of $(M_{\beta}, M_{\gamma})=(2,6)$, and the full $\widetilde{\Delta}(384)$ flavor symmetry of neutrinos.


\subsubsection{$(M_{\beta},M_{\gamma})=(4,4)$ case}

Next, let us consider  the case of the instanton zero-modes with $(M_{\beta},M_{\gamma})=(4,4)$.
In this case, $\widetilde{S}$ transformation for the matrix elements $(Z_1,Z_2,Z_3)^T \equiv \mathbf{Z}^T$, $\rho_{m^{(4,4)}}$, is the same as $\rho_{m^{(3,5)}}$ and $\rho_{m^{(2,6)}}$,~i.e. $\rho_{m^{(4,4)}}(\widetilde{S}) = \rho_{m^{(3,5)}}(\widetilde{S}) = \rho_{m^{(2,6)}}(\widetilde{S})$.
However, $\widetilde{T}$ transformation is not closed in the elements $\mathbf{Z}^T = \mathbf{Z'}^T + \mathbf{Z''}^T$ but closed in each of $\mathbf{Z'}^T=(Z'_1,Z'_2,Z'_3)^T$ and $\mathbf{Z''}^T=(Z''_1,Z''_2,Z''_3)^T$.
Their unitary matrices of $T$ transformation, $\rho_{m^{(4,4)'}}(\widetilde{T})$ and $\rho_{m^{(4,4)''}}(\widetilde{T})$, are given as
\begin{align}
\begin{array}{ll}
\rho_{m^{(4,4)'}}(\widetilde{T}) = e^{\pi i/4}
\begin{pmatrix}
1 & 0 & 0 \\
0 & e^{-\pi i/4} & 0 \\
0 & 0 & -1
\end{pmatrix}, & \rho_{m^{(4,4)'}}(\widetilde{T})^8 = \mathbb{I}, \\
\rho_{m^{(4,4)''}}(\widetilde{T}) =
\begin{pmatrix}
1 & 0 & 0 \\
0 & i & 0 \\
0 & 0 & -1
\end{pmatrix}, & \rho_{m^{(4,4)''}}(\widetilde{T})^4 = \mathbb{I},
\end{array}
\end{align} 
respectively.

On the other hand, we obtain
\begin{align}
\begin{array}{l}
{\rm det}[\rho_{inst^{(4,4)}}(\widetilde{S})^{(B,B)(C,C)}] = {\rm det}[\rho_{inst^{(4,4)}}(\widetilde{S})^{(C,C)(B,B)}] = {\rm det}[\rho_{inst^{(4,4)}}(\widetilde{S})^{(D,D)(D,D)}] = 1, \\
{\rm det}[\rho_{inst^{(4,4)}}(\widetilde{T})^{(B,B)(D,D)}] = {\rm det}[\rho_{inst^{(3,5)}}(\widetilde{T})^{(D,D)(B,B)}] = e^{5\pi i/4}, \\
{\rm det}[\rho_{inst^{(4,4)}}(\widetilde{T})^{(C,C)(C,C)}] = e^{-\pi i/2},
\end{array}
\end{align}
from Eq.~(\ref{eq:JrhoST}), and then we find
\begin{align}
\begin{array}{l}
{\rm det}[\rho_{inst^{(4,4)}}(\widetilde{S})^{(total)(total)}] = 1, \\
{\rm det}[\rho_{inst^{(4,4)'}}(\widetilde{T})^{(\mathbf{Z'})(\mathbf{Z'})}] = e^{5\pi i/4}, \  {\rm det}[\rho_{inst^{(4,4)''}}(\widetilde{T})^{(\mathbf{Z''})(\mathbf{Z''})}] = e^{-\pi i/2}.
\end{array}
\end{align}
This is the reason why the elements of $M_{ab}^{(4,4)}$, $\mathbf{Z}^T$, are closed under $\widetilde{S}$ transformation ,while $\widetilde{T}$ transformation is not closed in the elements $\mathbf{Z}^T$ but closed in each of $\mathbf{Z'}^T$ and $\mathbf{Z''}^T$.
Thus, in this case, the Majorana mass term generated by instanton zero-modes with $(M_{\beta}, M_{\gamma})=(4,4)$, $M_{ab}^{(4,4)}(\tau) N_a^{(8;0,0;1)}(x) N_b^{(8;0,0;1)}(x)$ is invariant under $\widetilde{S}$ and $\widetilde{T}^8$ transformation, while $\widetilde{T}$ transformation fully breaks this term unless $\widetilde{T}^{8n}$ for $\forall n \in \mathbb{Z}$.
Here, since $\widetilde{S}$ and $\widetilde{T}^{8}$ transformations for neutrinos satisfy
\begin{align}
\begin{array}{l}
\rho_N(\widetilde{S})^2 = -i\mathbb{I}, \\
\rho_N(\widetilde{S})^4 = \left[ \rho_N(\widetilde{S})\rho_N(\widetilde{T}^8) \right]^4 = -\mathbb{I}, \\
\rho_N(\widetilde{S})^8 = \left[ \rho_N(\widetilde{S})\rho_N(\widetilde{T}^8) \right]^8 = \rho_N(\widetilde{T}^8)^2 = \mathbb{I},
\end{array}
\end{align}
 they are the unitary representation of $\widetilde{\Sigma}(8) \equiv (Z_2 \times Z_2) \rtimes Z_8$~\footnote{It becomes the quadruple covering group of $\Sigma(8) \simeq (Z_2 \times Z_2) \rtimes Z_2$.}, where the generators of $Z_2$, $Z'_2$, and $Z_8$ are given by
\begin{align}
\begin{array}{l}
\rho_N(p) = \rho_N(\widetilde{T}^8), \\
\rho_N(q) = \rho_N(\widetilde{S}) \rho_N(\widetilde{T}^8)^2 \rho_N(\widetilde{S})^{-1}, \\
\rho_N(r) = \rho_N(\widetilde{S}),
\end{array}
\end{align}
respectively, and they satisfy
\begin{align}
\begin{array}{l}
\rho_N(p)^2 = \rho_N(q)^2 = \rho_N(r)^8, \\
\rho_N(p) \rho_N(q) = \rho_N(q) \rho_N(p), \ \rho_N(r)\rho_N(p)\rho_N(r)^{-1} = \rho_N(q), \ \rho_N(r)\rho_N(q)\rho_N(r)^{-1} = \rho_N(p).
\end{array}
\end{align}
Therefore, there remains $\widetilde{\Sigma}(8)$ flavor symmetry 
among the neutrino flavor symmetry $\tilde \Delta(384)$.

This result may be obvious because the instanton zero-modes, $\beta_1 \beta_2 \gamma_1 \gamma_2$ in the integral 
Eqs,~(\ref{eq:neutrinomass}), and (\ref{eq:beta-gamma-trans}) 
are not consistent with the modular symmetry.







\subsection{Three generations of  neutrinos with $(M; \alpha_1, \alpha_{\tau}; m)_N=(7;1/2,1/2;0)$}

The possible instanton zero-modes satisfying the condition (\ref{eq:zero-mode-condition}) 
are $(\beta, \gamma)^T$ with, 
\begin{align}
\begin{pmatrix}
(M; \alpha_1, \alpha_{\tau}; m)_{\beta} \\ (M; \alpha_1, \alpha_{\tau}; m)_{\gamma}
\end{pmatrix}
=
\begin{pmatrix}
(2;0,0;0) \\ (5;1/2,1/2;0)
\end{pmatrix},
\begin{pmatrix}
(3;1/2,0;0) \\ (4;0,1/2;0)
\end{pmatrix},
\begin{pmatrix}
(3;0,1/2;0) \\ (4;1/2,0;0)
\end{pmatrix},
\begin{pmatrix}
(3;0,0;0) \\ (4;1/2,1/2;0)
\end{pmatrix}. \notag
\end{align}
Then, the matrix can be written as
\begin{align}
M_{ab}
&= M_{ab}^{(2,5)} + M_{ab}^{(3,4)}. \label{eq:M7total} 
\end{align}
The first term $M_{ab}^{(2,5)}$ is the contribution due to the instanton zero-mode consistent with 
the modular symmetry and it has been obtained in the previous section.
The second term $M_{ab}^{(3,4)}$ includes the contributes due to the instanton zero-modes, which 
transform to others with different boundary conditions under the modular symmetry.
Similar to the previous case with $(M; \alpha_1, \alpha_{\tau}; m)_N=(8;0,0;1)$, 
we can compute $M_{ab}^{(3,4)}$ and investigate its flavor symmetry.
As a result, when we include $M_{ab}^{(3,4)}$,
only $Z_7 \times Z_4$ symmetry, whose generators are ${T'}^2$ in Eq.~(\ref{eq:PSL2Z7}) and $c^2$ in Eq.~(\ref{eq:PSL2Z7centerZ8}), respectively, remains among the neutrino flavor symmetry $PSL(2,Z_7) \times Z_8$.

\subsection{Comment}

As we have studied in the previous section, 
the D-brane instanton with zero-modes whose boundary conditions are consistent with the modular symmetry
breaks a single $Z_N$ subgroup of the modular flavor symmetry, and a certain normal subgroup is unbroken.
On the other hand, the D-brane instantons with zero-modes,
which have the boundary condition inconsistent the modular symmetry,
can violate the modular symmetry more severely.
That is, for the flavor symmetry $G \rtimes Z_N$, the above instantons break not only $Z_N$ but also $G$ to a smaller group, although $G$ may be anomaly-free through the discussion in section \ref{sec:modular-symm-anomaly}.
That may be obvious because 
the zero-modes, $\beta$ and $\gamma$, in Eq.~(\ref{eq:neutrinomass}) are not linear representations of the modular group in fixed boundary conditions
but they transform into the wavefunctions with different boundary conditions,
as shown in Eq.~(\ref{eq:beta-gamma-trans}).
One exception would be the neutrino mass for $(M; \alpha_1, \alpha_{\tau}; m)_N=(8;0,0;1)$ 
due to the instanton with $(M_\beta,M_\gamma) = (3,5)$, i.e., $M^{(3,5)}_{ab}$.
It includes the instanton zero-modes which are inconsistent with the modular symmetry in terms of the boundary condition. 
However, the full $\tilde \Delta(384)$ remains.
Its reason is unclear.
Its reason may be the summation over all of four SS phases,  
while the other cases are partial summation over SS phases.
Alternatively, this result may be just accidental.

The condition  (\ref{eq:zero-mode-condition}) does not prohibit the appearance of instanton zero-modes
which transform into others with different boundary conditions under the modular symmetry.
However, if we require the consistency with the modular symmetry, 
such instanton zero-modes would be forbidden.
It is unclear whether such instanton zero-modes can appear or not.
It may be concerned with the consistency condition with the string theory.
Throughout this paper, we have investigated the low energy field theory aspects of the D-brane models.
We have not severely taken into account of the stringy consistency conditions of D-brane instantons 
such as the number of the neutral zero modes.
They may restrict the possible configurations of D-brane instantons which can contribute to the superpotential,
and a part of the non-perurbative Majorana mass matrices proposed in Section \ref{sec:Modularflavoranomaly} and \ref{sec:more}
would be prohibited in full stringy models, and anomaly free part of the modular symmetry may be recovered.
Investigating full string model is interesting, but it is beyond our scope in the present work.
We would study this issue more elsewhere from the viewpoint of string theory.


\section{Conclusion}
\label{sec:conclusion}

We have studied the modular symmetry anomaly in magnetized orbifold models.
Non-perturbative effects can break the tree level symmetry.
The neutrino mass terms are important in particle physics, and 
can be induced by D-brane instanton effects.
They can break the modular symmetry.
Thus, we have studied the modular flavor symmetry anomalies of Majorana neutrino mass terms in 
concrete models with three generations of neutrinos on magnetized $T^2/Z_2$ orbifold explicitly.

It is found that the modular weight of neutrino mass terms does not match with other terms in 
tree-level Lagrangian.
The sum over weights of the instanton zero-modes $\beta_1 \beta_2 \gamma_1 \gamma_2$ is the 
origin of this difference.
This has significant meaning in model building of 4D modular flavor models although it would be cancelled by the shift of the axions through the generalized GS-mechanism.

In addition, the neutrino mass terms can break the tree-level modular flavor symmetry 
to its normal subgroup and a single $Z_N$ symmetry is broken, 
although there is an example that the full tree-level flavor symmetry remains.
This point is also important  in model building of 4D modular flavor models.
Among the tree-level flavor symmetry in the Yukawa coupling terms, 
its anomaly-free subgroup may remain in the neutrino mass terms.

When we include the effects due to the instanton zero-modes, which 
transform ones with different boundary conditions, 
the neutirno mass terms break the flavor symmetry to much smaller group.
It is unclear whether we must include such effects or not.
That is beyond our scope, and 
we would study this issue more elsewhere from the viewpoint of string theory.

It is also important to extend our analysis to active neutrino mass terms through the see-saw mechanism by combining Yukawa couplings as well as charged lepton mass terms.
We will study elsewhere those lepton mass terms, related to their masses and lepton flavor mixing, in terms of the modular symmetry\footnote{As for quark sector, their flavor structure such as their masses and flavor mixing is compatible with texture structure and a kind of texture structure can be obtained by considering the modular symmetry at fixed points for the modulus $\tau$~\cite{Kikuchi:2021yog}.}.

\vspace{1.5 cm}
\noindent
{\large\bf Acknowledgement}\\

This work was supported by JST SPRING Grant Number JPMJSP2119(SK) and 
JSPS KAKENHI Grant Number JP20J20388(HU).




\end{document}